\documentclass[fleqn,usenatbib,useAMS]{mnras}

\DeclareRobustCommand{\VAN}[3]{#2}
\let\VANthebibliography\thebibliography
\def\thebibliography{\DeclareRobustCommand{\VAN}[3]{##3}\VANthebibliography}

\usepackage{graphicx}	
\usepackage{amsmath}	
\usepackage{amssymb}	
\usepackage{multicol}        
\usepackage[multiple]{footmisc}
\usepackage{bm}		
\usepackage{pdflscape}	

\newcommand\xmm{{\it XMM-Newton}}

\newcommand\rxte{{\it RXTE}}
\newcommand\iue{{\it IUE}}
\newcommand\astrosat{{\it AstroSat}}

\newcommand\swift{{\it Swift}}

\usepackage{threeparttable}
\usepackage{array}
\newcolumntype{P}[1]{>{\centering\arraybackslash}p{#1}}
\usepackage[T1]{fontenc}
\usepackage{mathtools}
\usepackage{nccmath}
\usepackage[english]{babel}%
\usepackage{framed}
\usepackage[normalem]{ulem}
\usepackage{xcolor}
\usepackage{braket}
\usepackage{amsfonts}
\usepackage{enumerate}
\usepackage[utf8]{inputenc}
\usepackage{hyperref}
\usepackage{subcaption}
\usepackage{placeins}
\usepackage{floatrow}
\usepackage{booktabs,caption}
\usepackage{microtype}
\usepackage{color}
\usepackage{multicol}
\usepackage{multirow}
\usepackage{floatrow}
\usepackage{lipsum}
\usepackage{verbatim}
\usepackage{derivative}
\usepackage{epstopdf}
\usepackage{rotating}
\usepackage{fancybox}
\usepackage{pstricks}
\usepackage{pst-plot}
\usepackage{pstricks-add}
\usepackage{caption}
\usepackage{tabularx}

\title{Contrasting X-ray/UV time--lags in Seyfert 1 galaxies NGC~4593 and NGC~7469 using \textit{AstroSat} observations}

\author[K. Kumari et al.]{Kavita Kumari$^{1}$\thanks{E-mail: kavitak@iucaa.in},
G. C. Dewangan$^{1}$,
I. E. Papadakis$^{2,3}$, 
Max W. J. Beard$^{4}$, 
I. M. M\textsuperscript{c}Hardy$^{4}$,  
\newauthor
K. P. Singh$^{5}$,
D. Bhattacharya$^{1,6}$,
S. Bhattacharyya$^{7}$,
S. Chandra$^{8,9}$ 
\\
$^{1}$Inter-University Center for Astronomy and Astrophysics, Pune 411007, India\\
$^{2}$Department of Physics and Institute of Theoretical and Computational Physics, University of Crete, 71003 Heraklion, Greece\\
$^{3}$Institute of Astrophysics, FORTH, GR-71110 Heraklion, Greece\\
$^{4}$University of Southampton, University Road, Southampton SO17 1BJ, UK\\
$^{5}$IISER Mohali, Knowledge City, Sector 81, Manauli PO, SAS Nagar, Punjab 140306, India\\
$^{6}$Ashoka University, National Capital Region P.O. Rai, Sonepat Haryana-131029, India
\\
$^{7}$Department of Astronomy and Astrophysics, Tata Institute of Fundamental Research, Homi Bhabha Road, Mumbai 400005, India 
\\
$^{8}$South African Astronomical Observatory, Cape Town, South Africa\\
$^{9}$Center for Space Research, North-West University, Potchefstroom, South Africa 
}

\date{Accepted XXX. Received YYY; in original form ZZZ}

\pubyear{2022}

\begin{document}
\label{firstpage}
\pagerange{\pageref{firstpage}--\pageref{lastpage}}
\maketitle{}

\begin{abstract}
 
We study accretion disk-corona connection in Seyfert 1 galaxies using simultaneous UV/X--ray observations of NGC~4593 (July 14--18, 2016) and NGC~7469 (October 15--19, 2017) performed with  \astrosat{}. We use the X--ray (0.5--7.0 keV) data acquired with the Soft X-ray Telescope (SXT) and the  UV (FUV: 130--180 nm, NUV: 200--300 nm) data obtained with the Ultra-Violet Imaging Telescope (UVIT). We also use the contemporaneous \swift{} observations of NGC~4593 and demonstrate \astrosat{}'s capability for X--ray/UV correlation studies. We performed UV/X-ray cross-correlation analysis using the Interpolated and the Discrete Cross-Correlation Functions and found similar results. In the case of NGC~4593, we found that the variations in the  X-rays lead to those in the FUV and NUV bands by $\sim 38{\rm~ks}$ and $\sim 44{\rm~ks}$, respectively. These UV lags favour the disk reprocessing model, they are consistent with the previous results within uncertainties. In contrast, we found an opposite trend in NGC~7469 where the soft X-ray variations lag  those in the FUV and NUV bands by $\sim 41{\rm~ks}$ and $\sim 49{\rm~ks}$, respectively. The hard lags in NGC~7469  favour the Thermal Comptonization model.  Our results may provide direct observational evidence for the variable intrinsic UV emission from the accretion disk which acts as the seed for thermal Comptonization in a hot corona in a lamp-post like geometry. The non-detection of disk reverberation photons in NGC~7469, using \astrosat{} data, is most likely due to a high accretion rate resulting in a hot accretion disk and large  intrinsic emission.

\end{abstract}




\section{Introduction}
Active Galactic Nuclei (AGN) are one of the most luminous objects in the universe powered by  accreting supermassive black holes at the centres of the galaxies \citep{lynden1969galactic}. AGN exhibits complex spectra covering almost the full range of electromagnetic bands from radio to X-rays.
According to the current widely accepted models, the inner part of the optically thick and geometrically thin accretion disk emits thermal photons in the optical/UV band and sometimes in the soft X-ray band depending on the black hole mass ($M$) and the accretion rate ($\dot{m}$) relative to the Eddington rate, $T\propto \left( \frac{\Dot{m}}{M}\right)^{\frac{1}{4}}$  \citep{shakura1973black}. These thermal photons can be up-scattered via inverse Compton scattering off the free electrons in the optically thin, hot corona ($kT_e \sim$ 100 keV). This process is known as Thermal Comptonization  and generates the primary \textit{``power-law component"} with a high energy cut-off in the X-ray band \citep{sunyaev1980comptonization, haardt1993anisotropic}. The exact geometry and the physical mechanisms responsible for the origin of the corona are still murky \citep{poutanen1996two}. A fraction of the power-law component irradiates the disk where some of the photons get absorbed and some get reflected which we detect as the \textit{``reflected component"} (e.g. \citealt{nandra1994ginga}). This reflected component (Fe K$\alpha$ line and Compton hump) evinces the signature of various interesting interactions between X-rays and accretion disk material. The  photons that get reflected  from the inner disk also exhibit the signature of relativistic effects such as gravitational redshift, light bending and Doppler boosting  due to the strong gravitational field of the central black hole  (see, e.g. \citealt{fabian1989x, laor1991line, miniutti2004light, dovvciak2004extended, dauser2012spectral} ). The X-ray photons which are not reflected are absorbed and contribute to the disk heating, 
hence increasing UV/optical emission from the disc. Since X-rays are variable in most AGN, X--ray illumination of the disc predicts
that the disc emission should also be variable. The study of correlations between X-rays and UV/optical variations in AGN can in principle provide valuable information regarding the disc-corona geometry in AGN. This possibility has led to a tremendous observational effort in the last 10-15 years (see e.g. \citealt{mchardy2014swift, mchardy2018x, edelson2019first, hernandez2020intensive, kara2021agn}) as
well as a detailed theoretical study of the expected correlations in the case of the X--ray lamp post model (i.e. \citealt{kammoun2019hard, kammoun2021modelling,  kammoun2021uv}).


\par The study of temporal variability and their correlation across different energy bands can provide essential clues on the connection between disk and coronal emission and the possible physical mechanism responsible for the observed variability. If the variable part of the UV is dominated by the X-ray reprocessing from the disk, then we expect X-ray photons to lead over the UV photons \citep{krolik1991ultraviolet}. Indeed, a number of Seyfert 1 AGNs exhibit UV lags that are generally consistent with X-ray reprocessing \citep{kammoun2021modelling}.
In addition, if the Compton up-scattering of the UV seed photons emitted from the accretion disk is the dominant process for the production of the X-rays, then we expect X-rays to lag behind UV photons and detection of such time--lags would strongly support the Thermal Comptonization model (e.g. \citealt{adegoke2019uv, nandra1998new, arevalo2005x}). We should keep in mind that, the Thermal Comptonization of the disk photons is believed to operate all the radio-quiet AGNs while the disk reprocessing signals actually depend on the corona/disk configuration (or geometry). For a simple lamp-post geometry, we expect both signals to be detected, but X-ray/UV lead or lag is sensitive to the dominant processes responsible for the observed variability in their light curves. 

Also, the measured time--lag can put important constraints on the size of the disk and possible underlying physical mechanism for the emission.  Lags of a few days can be due to  light crossing time between the disk and the corona (lamp-post model) which is possible for both scenarios i.e. reprocessing or Thermal Comptonization. The case where X-rays lag behind UV photons by  $\sim years$ can be explained by the propagating-fluctuation model \citep{lyubarskii1997flicker, arevalo2006investigating}, in which the fluctuations in the accretion flow propagates inward in the accretion disk. Some AGNs exhibit correlated X-ray/UV emission but without significant delays (e.g. \citealt{breedt2009long}). 
The different lag behaviour obtained from the studies of various sources indicates that the X-ray/UV emission mechanism is complex and the study of more AGNs using well-sampled, good quality simultaneous X-ray/UV data is required. 

\par With its UV and X-ray payloads, the Indian multi-wavelength astronomy mission \astrosat{} is well suited for simultaneous X-ray/UV observations. We have performed long X-ray/UV observations of bright Seyfert 1 galaxies, with \astrosat{}. Here we demonstrate \astrosat{}'s capability for correlation studies, and measure time--lags between X-ray/UV variations in two AGN,  NGC~4593 and NGC~7469.

NGC~4593 is a barred galaxy at a redshift of $z=0.009$ \citep{strauss1992redshift}, it hosts a Seyfert 1 AGN with  a black hole mass in the range of $(1-10) \times 10^6 M_{\odot}$ \citep{peterson2004central, denney2006mass}. The AGN has been found to be highly variable in optical, UV, and X-ray bands \citep{ursini2016high}. Earlier, this source was observed by \swift{} for 22 days in July--August 2016, and these data were studied by \cite{mchardy2018x}. They found  the X-ray variations  to lead the variations in UVW2 band  by $\sim 0.66 \pm 0.15$ days ($\sim$ 57 ks). 
\cite{pal2017correlated} have also studied the cross-correlation and found time--lag using the \swift{}/UVOT and \swift{}/XRT data of NGC~4593 observed from  13 July 2016 to  5 August 2016. They reported that the observed variability in the optical/UV band light curve lags the variations in the X-ray band light curve on a short timescale of a few days ( $\sim 0.4-1.5$ days) and strongly supports the disk reprocessing model.

NGC~7469 is a type I Seyfert galaxy located at a redshift of $z = 0.016$, it has  a central SMBH of mass $(1-6) \times 10^7 M_{\odot}$ \citep{peterson2014reverberation, shapovalova2017long, nguyen2021black}. Previously,  \cite{nandra1998new} used the 30~days long \rxte{} data acquired in June--July 1996 
and  46~days long \iue{} data acquired simultaneously with the \rxte{} data and found the largest positive correlation for the UV (1315~\AA) to lead the X-rays by $\sim 4$ days. Their results favoured the Thermal comptonization model. Using the same \iue{}/\rxte{} data, \cite{petrucci2004physical} studied  correlations between X-ray spectral parameters and X-ray/UV fluxes. One of the crucial results obtained by \cite{petrucci2004physical} is the anti-correlation between the UV flux and the coronal temperature which can be explained as the cooling of the hot corona with increasing UV flux, as predicted in the Thermal Comptonization model.
%
\cite{pahari2020evidence} has also analysed the same data set and found that the UV continuum light curve leads the 2--10 keV X-ray light curve by $3.49 \pm 0.22$ days. However, if variations slower than 5~days are filtered out from the X-ray light curve, the UV variability lags the X-ray variability by $0.37 \pm 0.14 $ days. Such soft lag is consistent with the reprocessing of X-rays in the accretion disk into the UV emission.

In this paper, we derive cross-correlation and lag between  X-ray and UV variations in NGC~4593 and NGC~7469 using  \astrosat{} observations and study the interplay of photons between the disk and the corona. In section~\ref{observation}, we describe our observations and data reduction procedure. We study  X-ray/UV variability  in section~\ref{sec:fvar} followed by the cross-correlation analysis of NGC~4593 and NGC~7469 in sections~\ref{section_ngc4593} and \ref{section ngc7469}, respectively. We discuss our results in section~\ref{discussion}, followed by conclusions in section~\ref{conclusion}.

\begin{figure*}[!ht]

\centering
\begin{subfigure}{0.3\textwidth}
\includegraphics[width=1\textwidth]{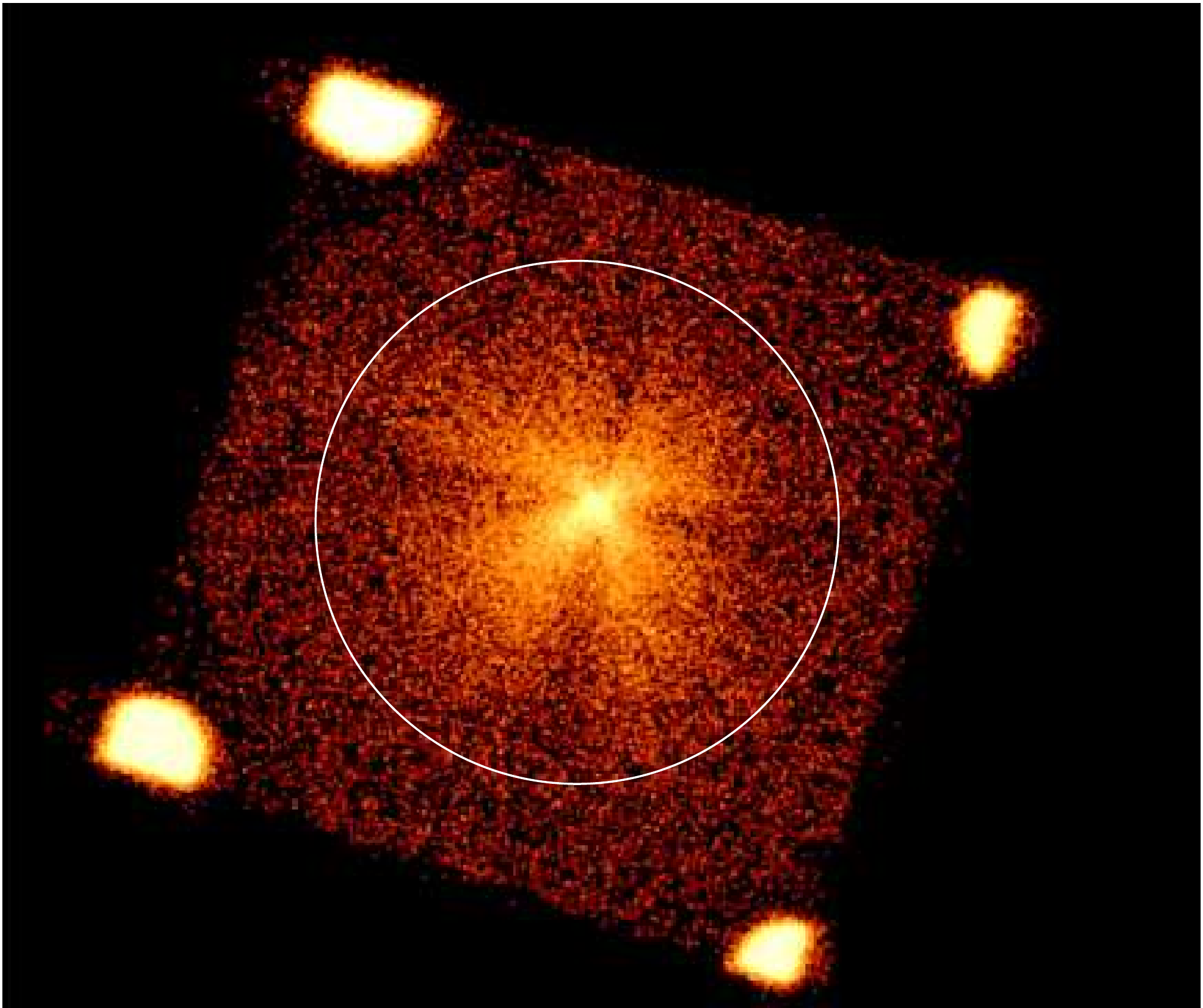}
\caption{}
\label{sxt_ngc4593}
\end{subfigure}
\hfill
\begin{subfigure}{0.3\textwidth}
\includegraphics[trim=20 0 25 0,clip,width=1\textwidth]{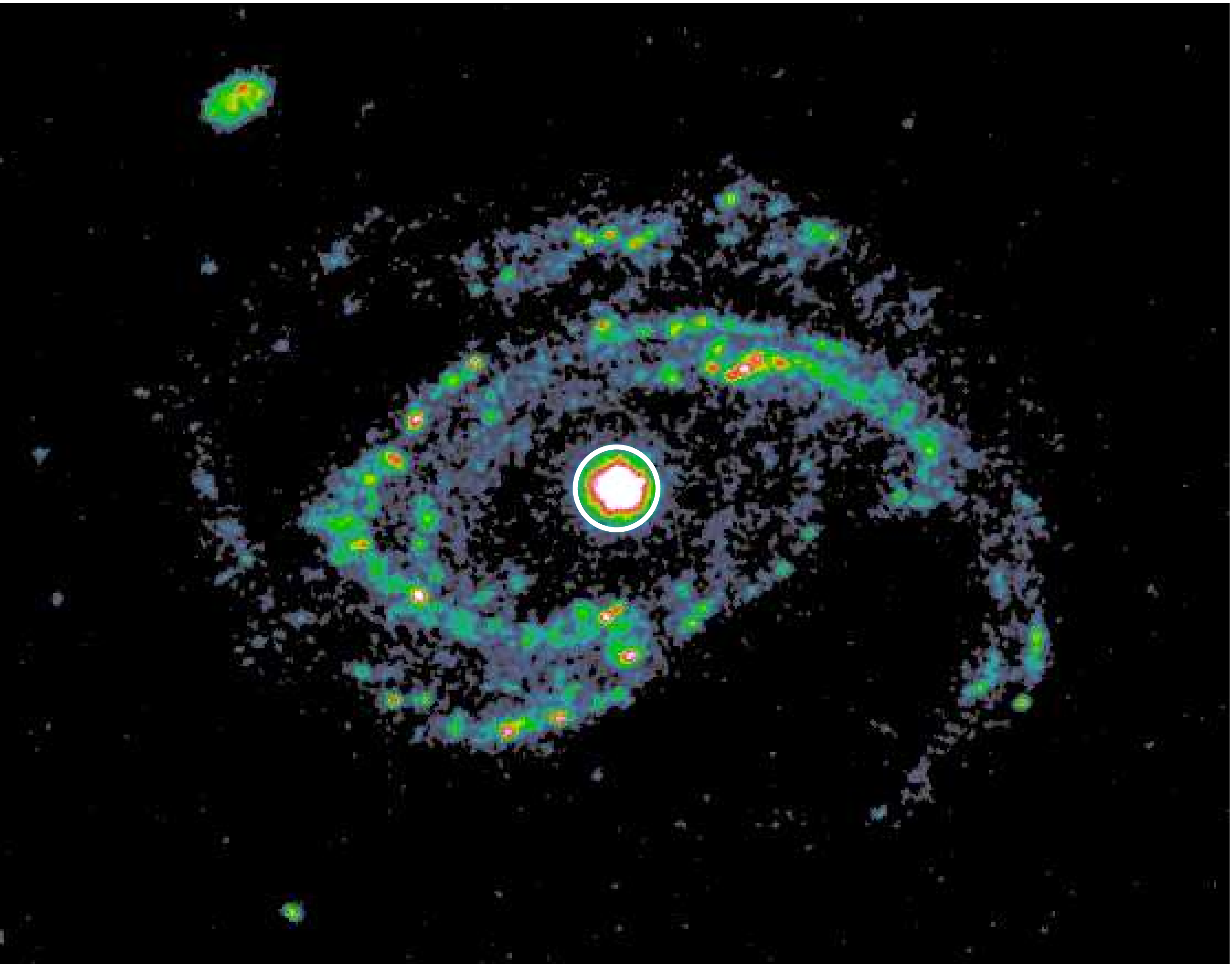}
\caption{}
\label{fuv_ngc4593}
\end{subfigure}
\hfill
\begin{subfigure}{0.3\textwidth}
\includegraphics[trim=0 0 0 8,clip,width=1\textwidth]{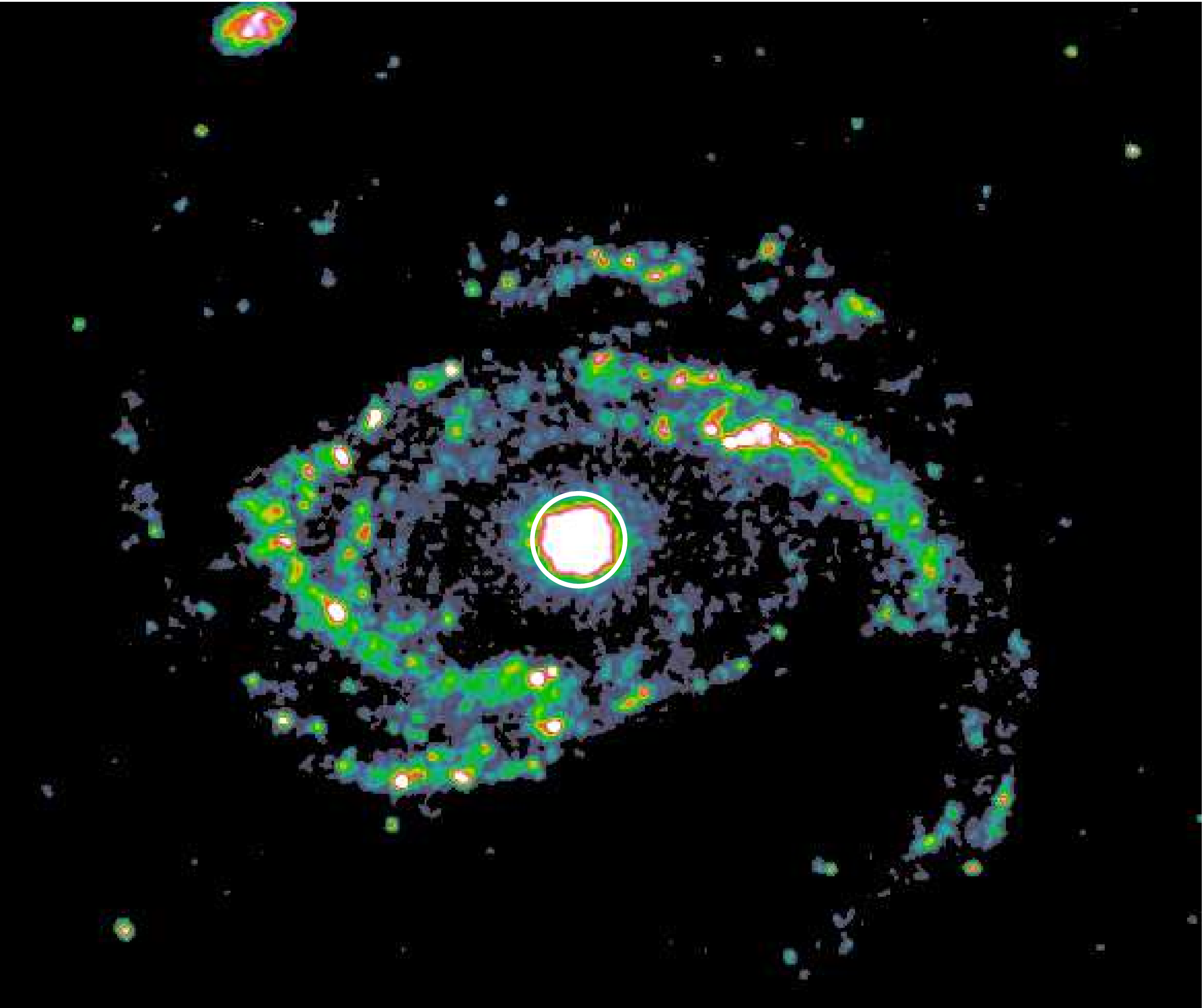}
\caption{}
\label{nuv_ngc4593}
\end{subfigure}
\caption{Images of NGC~4593 in different energy bands: (a) SXT (0.5 -- 7.0 keV) (b) FUV/BaF2 (F154W; $\lambda_{mean}=1541$~\AA) and (c) NUV/NUVB4 (N263M; $\lambda_{mean}=2632$~\AA). The circular regions (white) depict the source extraction regions of 15 arcmins for SXT and 0.173 arcmins (10.4 arcsecs) for UVIT images.} 

\label{images_n4593}
\end{figure*}

\begin{figure*}[!ht]

\centering
\begin{subfigure}{0.3\textwidth}
\includegraphics[width=1\linewidth]{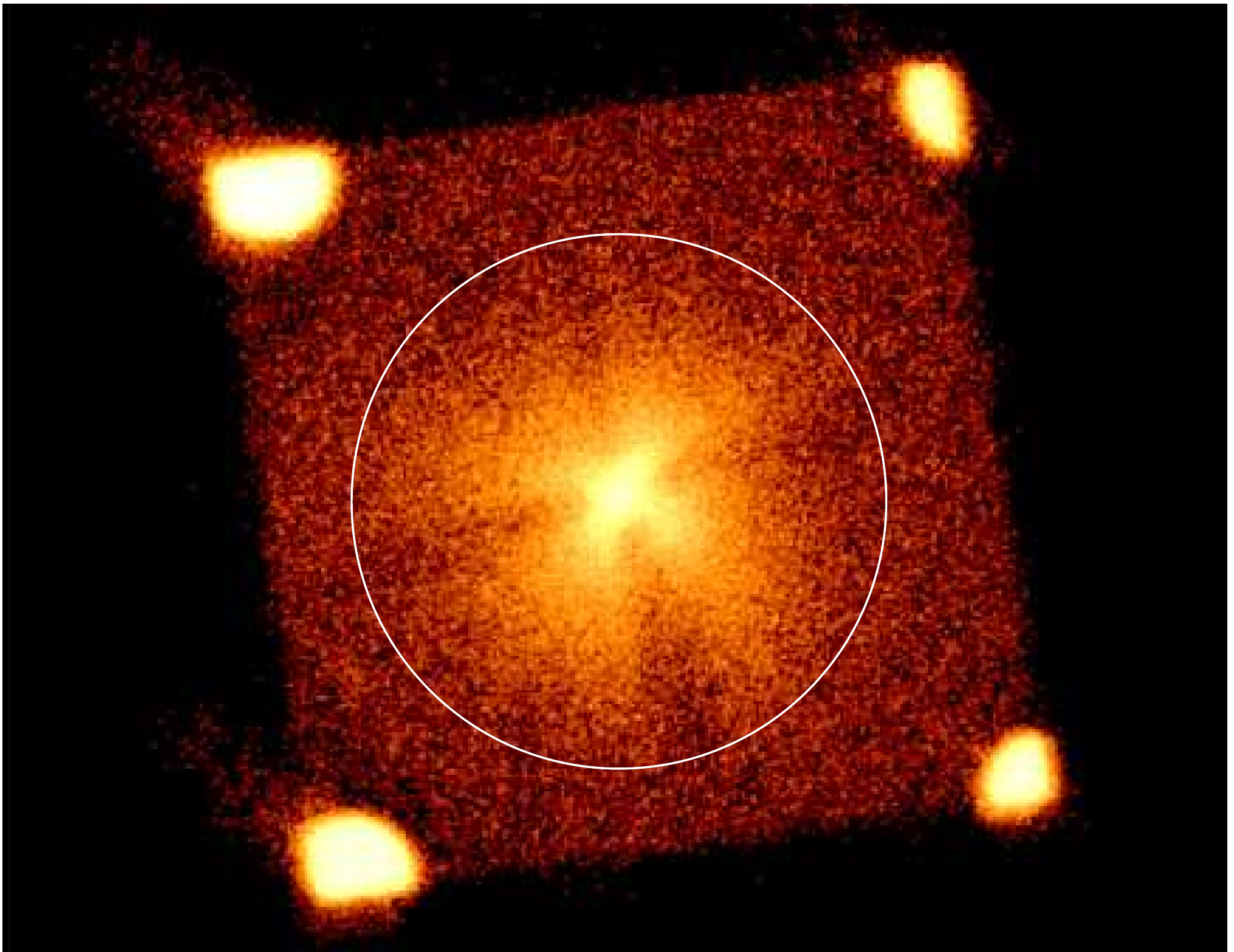}
\caption{}
\label{sxt_ngc7469}
\end{subfigure}
\hfill
\begin{subfigure}{0.3\textwidth}
\includegraphics[width=1\linewidth]{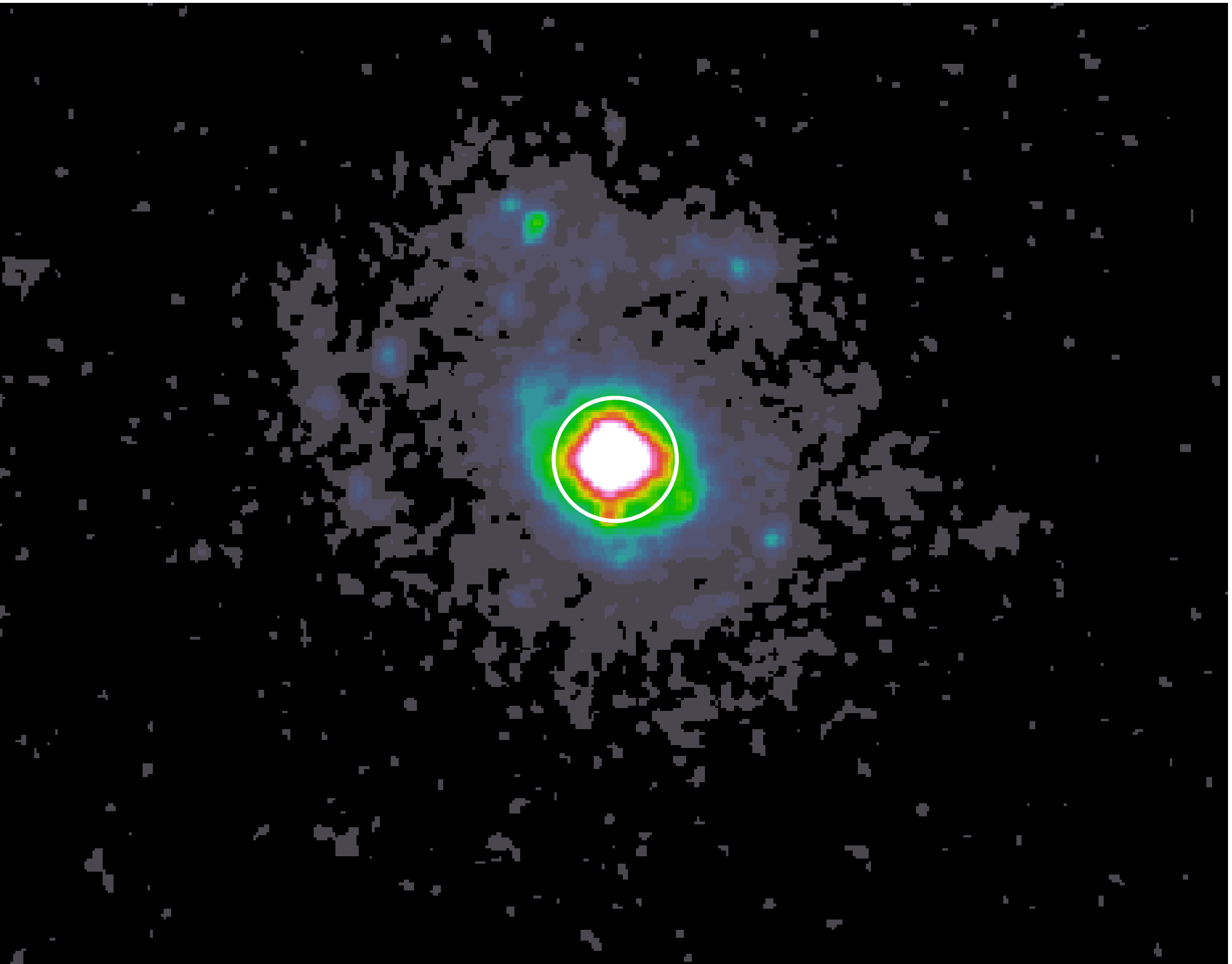}
\caption{}
\label{fuv_ngc7469}
\end{subfigure}
\hfill
\begin{subfigure}{0.3\textwidth}
\includegraphics[width=1\linewidth]{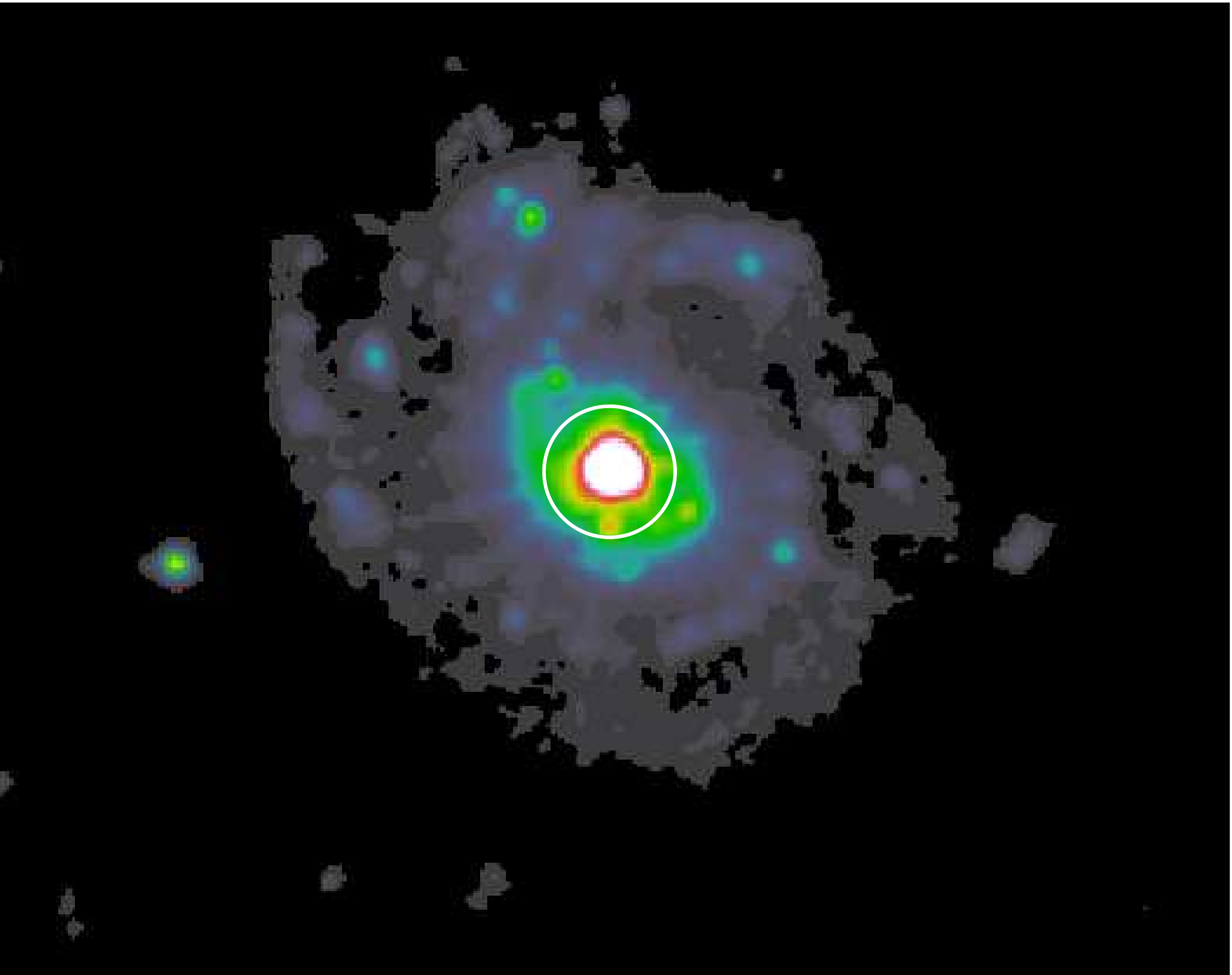}
\caption{}
\label{nuv_ngc7469}
\end{subfigure}
\caption{X-ray and UV images of NGC~7469: (a) SXT (0.5 -- 7.0 keV) (b) FUV/Silica (F172M; $\lambda_{mean}=1717$~\AA) and  (c) NUV/NUVB13 (N245M; $\lambda_{mean}=2447$~\AA). The white  circles depict the source extraction regions of 15 arcmins for SXT and 0.125 arcmins (7.5 arcsecs) for UVIT images. }
\label{images_n7469}
\end{figure*}

\begin{table*}[!ht]
\caption{Details of the \astrosat{} observations.}
\renewcommand{\arraystretch}{2}
\begin{tabular}{P{0.09\linewidth}P{0.20\linewidth}P{0.14\linewidth}P{0.18\linewidth}P{0.12\linewidth}P{0.13\linewidth} } \\
\hline
\hline
\textbf{Source} & \textbf{Observation ID} & \textbf{Obs. date} & 
\textbf{Energy band/Filter} & \textbf{Mean count rate (c/s)} &\textbf{Net exposure time (ks)}\\
\hline

NGC~4593 & G05\_219T01\_9000000540  & 14-18 July 2016 & FUV/F154W & 4.48   &  16.3   \\

& & & NUV/N263M & 17.95  &17.1     \\

& & & SXT/(0.5-7.0 keV) & 0.58  & 45.5  \\

NGC~7469 & G08\_071T02\_9000001620  & 15-19 Oct. 2017 & FUV/F172M & 4.53 & 45.1 \\

& & & NUV/N245M & 40.32  &  53.0     \\

& & & SXT/(0.5-7.0 keV) & 0.75  &  108.3    \\

\hline
\hline
\end{tabular}
\label{data}
\end{table*}

\section{Observations and Data Reduction}\label{observation}
\subsection{\astrosat{} observations}
We utilised the multi-wavelength capability of the \astrosat{} mission that was launched on 30~September 2015, and is being operated by the Indian Space Research Organisation (ISRO). \astrosat{} carries four co-aligned scientific payloads -- the Ultra-Violet Imaging Telescope (UVIT; \citealt{tandon2017orbit, tandon2020additional}), the Soft X-ray Telescope (SXT; \citealt{singh2016orbit, singh2017soft}), the Large Area Proportional Counters (LAXPC; \citealt{agrawal2017large,antia2017calibration}) and the Cadmium Zinc Telluride Imager (CZTI;  \citealt{vadawale2016orbit, bhalerao2017cadmium}). Here, we used the data from the UVIT and SXT. The UVIT consists of twin telescopes that image the sky in three different channels -- Far Ultra-Violet (FUV, 130--180 nm), Near Ultra-Violet (NUV, 200--300 nm) and Visible (VIS, 320--550 nm). The VIS channel is used for tracking purposes only, while the FUV and NUV channels equipped with multiple broadband filters and slitless gratings provide science images with a good spatial resolution (FWHM\footnote{FWHM: Full Width at Half Maximum}$\sim 1 - 1.5{\rm~arcsecs}$) and low-resolution spectra \citep{tandon2017orbit} within a circular field of view (FOV) of $\sim 28{\rm~arcmins}$ diameter. The FUV and NUV channels operate in the photon counting mode with sub-second timing capability. The SXT is an imaging spectrometer similar to the \swift{}/XRT but with inferior spatial resolution (FWHM$\sim 2{\rm~arcmins}$, half-power diameter $\sim 10{\rm~arcmins}$). This leaves no source-free area in the CCD detector, and observations of blank sky are used for background correction.

\astrosat{} observed NGC~4593 during 4--18 July 2016 with the CZTI as the primary instrument (Obs.~ID:~G05\_219T01\_9000000540). We operated the SXT in the photon counting (PC) mode using the entire sensitive area of the CCD camera. We used the broadband UVIT filters FUV/BaF2 (F154W; $\lambda_{mean}=1541$ \AA, $\Delta\lambda = 380$ \AA) and NUV/NUVB4 (N263M; $\lambda_{mean}=2632$ \AA, $\Delta\lambda = 275$ \AA) and acquired the data from the full field. 

Depending on the constraints, the observing efficiency of the instruments varies with time. The UVIT has the most stringent constraints that result in the lowest observing efficiency. We list the details of the observations including the net exposure time and count rates in Table~\ref{data}. The \astrosat{} observations of NGC~4593 were performed in coordination with \xmm{} and \swift{}, and the data obtained from these coordinated observations are studied in \cite{beard2022timescale}. The main purpose here is to use different cross-correlation techniques and establish \astrosat{}'s capability for X-ray/UV correlation studies by comparing with the exact simultaneous data acquired with the \swift{}. We also account for varying observing times in different \astrosat{} orbits in our cross-correlation analysis.

\astrosat{} observed NGC~7469  during 15--19 October 2017 with the SXT in the PC mode as the primary instrument (Obs. ID:~G08\_071T02\_9000001620; Table~\ref{data}) and  UVIT in the full field using the filters FUV/Silica (F172M; $\lambda_{mean}=1717$~\AA, $\Delta\lambda = 125$~\AA) and NUV/NUVB13 (N245M; $\lambda_{mean}=2447$~\AA, $\Delta\lambda = 280$~\AA).  Here, we did not use the LAXPC and CZTI data due to the high background.

\subsection{UVIT data reduction}\label{2.2}

We obtained the Level1 (L1) UVIT data from the \astrosat{} data archive\footnote{\url{https://astrobrowse.issdc.gov.in/astro_archive/archive/Home.jsp}}.   We used the \textsc{CCDLAB} UVIT Pipeline \citep{postma2017ccdlab} and processed the L1 data. The pipeline   corrects for field distortion, centroiding bias, and pointing drift and applies flat-fielding. It also accounts for frames rejected due to cosmic rays or missing from the L1 data. We generated the pointing drift series  using  the VIS images taken every second and corrected for the pointing drift in each orbit. We generated cleaned images and centroid lists for each orbit.
We then aligned the orbit-wise, drift-corrected centroid lists and created a merged centroid list and merged image for each filter. We further optimised the PSF (point spread function) of the UV images in a single band by using channel-self correction i.e., by using the variations in the source positions in the merged centroid lists in the same band. We show the final merged FUV/NUV band images of NGC~4593 in  Fig.~\ref{fuv_ngc4593} and \ref{nuv_ngc4593}.  We used the circular regions with radii 25~pixels or 10.4~arcsecs, shown in white in Fig.~\ref{fuv_ngc4593} and \ref{nuv_ngc4593}, centred at the source position to extract the source counts.  
In the case of NGC~7469, we used a circular region with a radius of 18 pixels or 7.5~arcsecs centred at the source position, as shown in Fig.~\ref{fuv_ngc7469} and \ref{nuv_ngc7469}, to extract the source counts.

We also used circular regions in the source-free areas  to extract the background counts which we used to correct for the background contributions. We corrected the source counts for the saturation effect following the prescription described in \cite{tandon2017orbit}. We constructed the net source light curves in different bands by extracting the source and background counts from each of the cleaned orbit-wise images and correcting for the saturation effect and background contribution. For the photometry and light curve generation as described above, we have developed tools in the Julia language, these tools are part of the {\sc UVITTools} package described in \cite{dewangan2021calibration}. Each data point in the UV light curves corresponds to each orbit-wise image.

It is clear from  Fig.~\ref{fuv_ngc4593}, \ref{nuv_ngc4593} that the AGN in NGC~4593 completely dominates the UV emission from the central regions with the negligible contribution of diffuse emission from the host galaxy. This, however, does not seem to be the case for NGC~7469 (see Fig.~\ref{fuv_ngc7469} and \ref{nuv_ngc7469}) where diffuse emission is clearly seen. To estimate the contribution of the diffuse emission from the host galaxy, we performed radial profile analysis of NGC~7469 in the FUV and NUV bands  (see Appendix \ref{sec:radprof} for details). We found that the diffuse emission contributes only $\sim 3.5\%$ and $5\%$ to the total FUV and NUV emission within the source extraction regions shown in Fig.~\ref{fuv_ngc7469} and \ref{nuv_ngc7469}. We did not make any corrections for the small contribution of the host galaxy.

\subsection{SXT data reduction }\label{2.3}

We downloaded the orbit-wise L1 SXT data on NGC~4593 and NGC~7469 from the \astrosat{} archive and processed using the  {\sc SXTPIPELINE} (Version:~1.4b) available at  the SXT website\footnote{\url{https://www.tifr.res.in/~astrosat_sxt/index.html}} and obtained the orbit-wise level2 data that include the cleaned event list for each orbit. We merged
the orbit-wise event lists  using the Julia package  {\textsc SXTMerger.jl}\footnote{\url{https://github.com/gulabd/SXTMerger.jl}}, also available at the SXT website. 
We used the HEASoft tool \textsc{XSELECT} (version 2.4g) to extract the images and light curves from the merged event lists. In  Fig.~\ref{sxt_ngc4593} and \ref{sxt_ngc7469}, we show the SXT images of NGC~4593 and NGC~7469 and the circular regions (15~arcmins radius) centred at the source positions that we used to extract light curves.

Since the X-ray extraction regions are large, we searched for nearby X-ray sources whose X-ray emission could contribute significantly to the observed count rates from AGNs. To this end, we searched the 4XMM-DR12 source catalogue \footnote{\url{http://xmm-catalog.irap.omp.eu}} \citep{webb2020xmm} for sources within a 15~arcmins radius. We found that the combined flux contribution from all the other sources to the total flux of NGC~7469 and NGC~4593 is $<5$\% and $<6$\%, respectively. This small contribution from other X-ray sources is unlikely to affect our results.

We extracted the source light curves in the 0.5--7.0~keV band with 2.3775~s (time resolution of SXT) bins.
The large PSF with extended wings and the presence of four corner calibration sources leave no source-free regions in the SXT CCD. Therefore, we did not extract background light curves from the same data. The large number of SXT blank-sky observations performed by the instrument team  show that SXT background is steady and small ($0.08{\rm~counts~s^{-1}}$). We estimated the background count rate from the blank sky PHA spectrum  (SkyBkg\_com-
b\_EL3p5\_Cl\_Rd16p0\_v01.pha) provided by the SXT POC. We found that the SXT background contributes only $\sim 13 \%$ and $\sim 10 \%$ to the total count rates for NGC~4593 and NGC~7469  within $15{\rm~arcmins}$ circular region and in the $0.5-7.0\rm~kev$ band.


\subsection{Binning of light curves}\label{2.4}

The Earth's occultation and the passage through the south Atlantic anomaly leave  gaps in the light curves of cosmic X-ray sources observed by low Earth orbit satellites, such as \astrosat{}. Additional observing constraints further reduce observing efficiencies of UVIT and SXT to $\sim 15\%$ and $25\%$, respectively.  The bin-sizes in the UV light curves, extracted in section~\ref{2.2}, represent the actual observation duration or the exposure time for each orbit-wise image which can range from $\sim 500 - 1500{\rm~s}$, and the corresponding mid-points are the bin-times in the light curves. The average gap between the two data points is $\sim 97{\rm~minutes}$. However, we can also notice some large gaps in the UV light curves (see Fig.~\ref{LC_ngc4593} and \ref{LC_ngc7469}) which are due to missing orbit-wise images.

In the extraction of X-ray light curves using the {\sc xselect} tool, the bin times are assigned at the midpoints of the bins. If the bins are not fully exposed, this can be an issue as the midpoints of the exposed part of the bins can be different than the midpoints of the full bins. Rebinning light curves with the {\sc lcurve} tool also result in a similar issue. In many cases the astronomical data are unevenly sampled, we, therefore, do not expect data points at all times within a particular time-bin. The average count rate in a particular time-bin should lie at the average of times when we actually detected the photons as determined by the Good Time Interval (GTI) of the observation. The nominal $25\%$ observing efficiency of SXT results in the on-source time of $\sim 1500{\rm~s}$, thus leaving about gaps for $\sim 4500{\rm~s}$ in an orbit.
We, therefore, rebinned the SXT light curves extracted with $2.3775{\rm~s}$ bins by computing the average count rate for each continuous stretch of data within one orbit as determined by the GTI  and assigning the bin-time as the midpoint of the continuous stretch.

\begin{figure}[!ht]
\includegraphics[width=1.05\columnwidth]{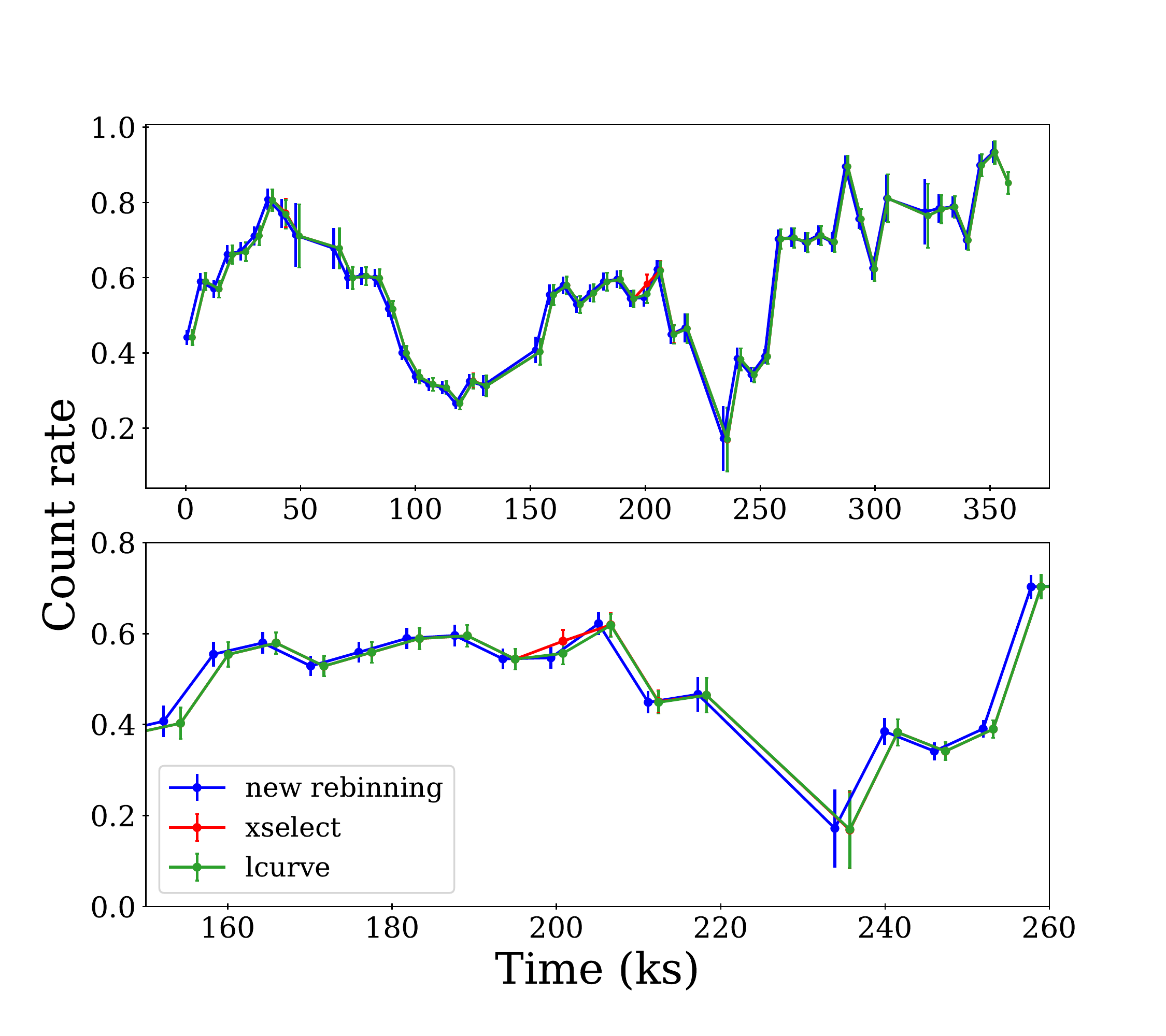}
\caption{\textit{Upper panel}- SXT light curves of NGC~4593 extracted using XSELECT (red), and \textit{lcurve} (green). The light curve rebinned by us manually is plotted in blue. \textit{Lower panel}- Same as above but zoomed in the time period between 150-260 ks for clear visibility of the shift in our light curve.}
\label{rebin_sxt}
\end{figure}

\begin{figure*}[!ht]
\centering
\begin{subfigure}{0.48\textwidth}
\includegraphics[width=1\linewidth]{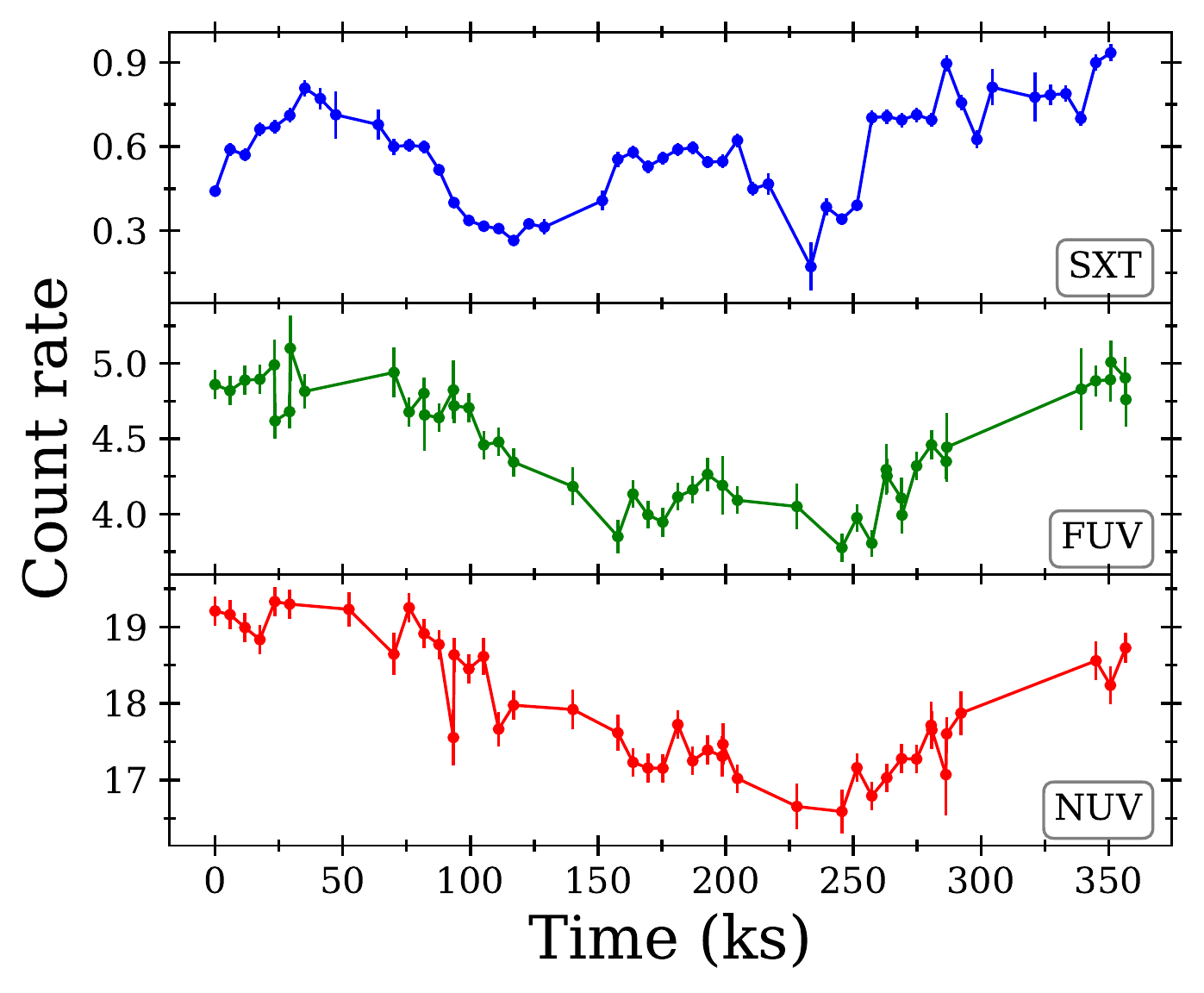}
\caption{}
\label{LC_ngc4593}
\end{subfigure}
\hfill
\begin{subfigure}{0.48\textwidth}
\includegraphics[width=1\linewidth]{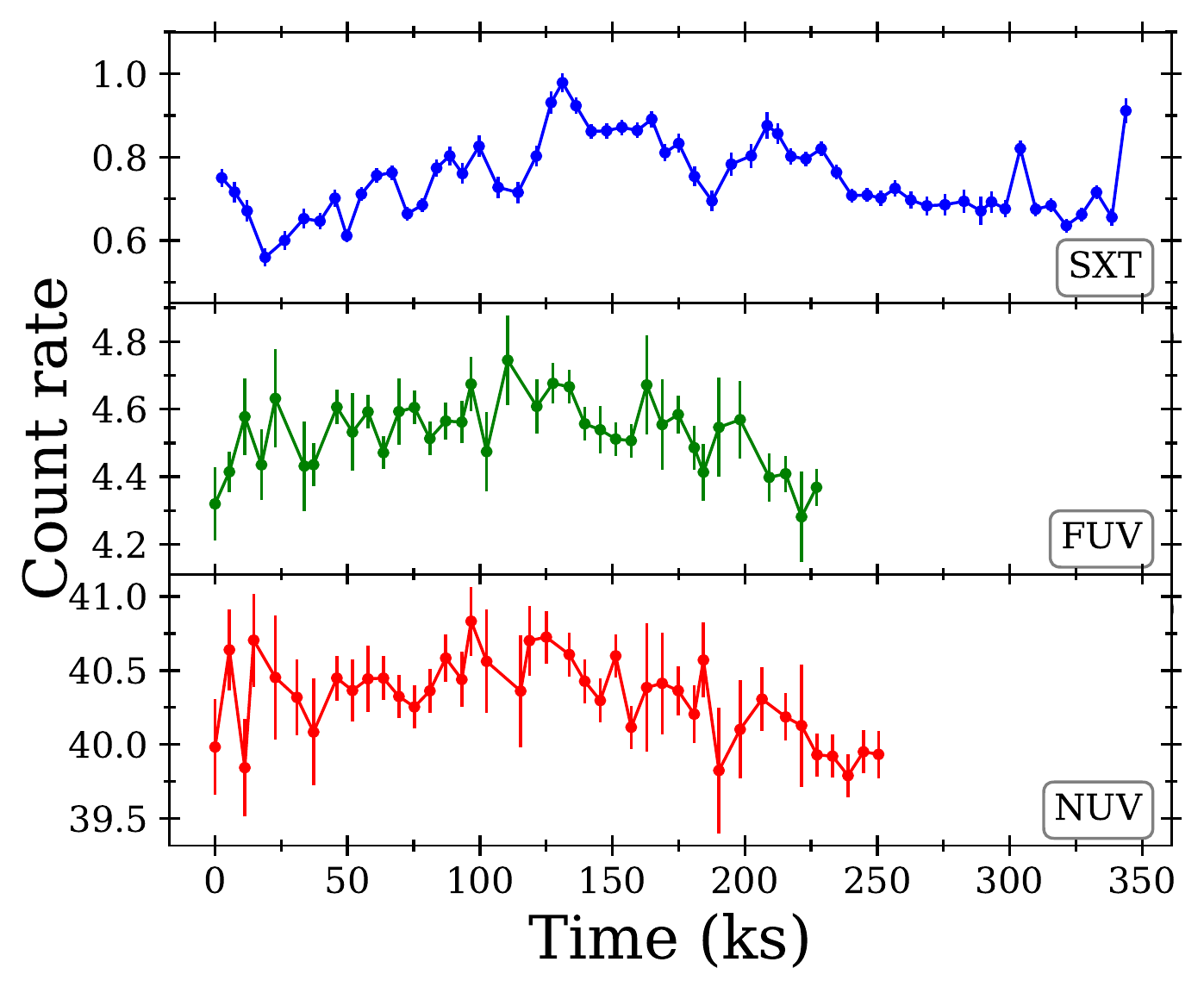}
\caption{}
\label{LC_ngc7469}
\end{subfigure}
\caption{\small{\astrosat{} light curves of (a) NGC~4593 in energy bands: SXT (0.5--7.0 keV), FUV/BaF2 ($\lambda_{mean}=$ 1541 \AA) and NUV/NUVB4 ($\lambda_{mean}=$ 2632 \AA)   (b) NGC~7469 in energy bands: SXT (0.5--7.0 keV), FUV/Silica ($\lambda_{mean}=$ 1717 \AA) and NUV/NUVB13 ($\lambda_{mean}=$ 2447 \AA). }}
\label{light_curves}
\end{figure*}

In Fig.~\ref{rebin_sxt}, we have plotted the light curves of NGC~4593 with 5820~s bins, obtained using {\sc xselect} (LC1) and \textit{lcurve} (LC2) and compare them with our new rebind light curve. We note that LC1 and LC2 overlap while our light curve is slightly shifted by a few ks. Sometimes even this minor shift in a light curve can give misleading estimates of X-ray/UV time--lag. We also rebinned the SXT light curve of NGC~7469 in a similar manner. 

\section{Light curves and their variability amplitude}
\label{sec:fvar}

In Fig.~\ref{LC_ngc4593} and \ref{LC_ngc7469}, we show the SXT, FUV and NUV light curves of NGC~4593 and NGC~7469, respectively. Both the sources are clearly variable in all bands,  and at all the sampled time-scales. Fast variability is more difficult to detect in the FUV and NUV light curves of NGC~7469, because the observations are of relatively small amplitude, and the errors due to the Poisson noise are comparable with the observed short term variations.

First, we tested whether the light curves shown in Fig.~\ref{LC_ngc4593} and \ref{LC_ngc7469} are variable. This is obviously the case with the X--ray light curves, but less so with the UV light curves, mainly of NGC~7469. To check the variability, we fitted a constant to the light curve and our results are as follows: $\chi^2=511/47$ degrees of freedom (dof), and $\chi^2=670/42$ dof, for the FUV and the NUV light curves of NGC~4593, respectively. The probability of the source being constant in these bands is $p_{nul}=2.7\times 10^{-79}$ and $p_{nul}=3.8\times 10^{-114}$, respectively. The respective results for NGC 7469 are as follows: $\chi^2=66.48/36$ dof, and $\chi^2=133/40$ dof, for the FUV and NUV light curves, while the respective $p_{nul}$ probabilities are: $1.5\times 10^{-3}$ and $7.4\times 10^{-12}$. These results indicate that both the FUV and the NUV light curves of this source are significantly variable.


As a first measure of the variability in the X-ray/UV bands, we calculated the fractional variability amplitude ($F_{var}$)  of the light curves. We used the formula as defined in \cite{vaughan2003characterizing},

\begin{equation}
    F_{var} = \sqrt{\frac{\sigma^2_{XS}}{\overline{x}^2}} = \sqrt{\frac{S^2 - \overline{\sigma^2_{err}}}{\overline{x}^2}},
\end{equation}

\noindent where  $\overline{x}$ and $S^2$ are the sample mean and variance of the light curves,  $\overline{\sigma^2_{err}}$ is the mean square error of the points in the light curve. Here, $\sigma^2_{XS}~=~S^2 - \overline{\sigma^2_{err}}$ is the excess variance which is an estimator of the intrinsic variance of the source. The error on $F_{var}$ is calculated using equation (B2) from \cite{vaughan2003characterizing}, i.e.,

\begin{equation}
   err (F_{var}) = \frac{1}{2F_{var}}err(\overline{\sigma^2_{NXS}}),
\end{equation}

\noindent where 

\begin{equation}
   err (\sigma^2_{NXS}) = \sqrt{\left(\sqrt{\frac{2}{N}}.\frac{\overline{\sigma^2_{err}}}{\overline{x}^2}\right)^2 + \left(\sqrt{\frac{\overline{\sigma^2_{err}}}{N}}. \frac{2F_{var}}{\overline{x}} \right)^2}.
\end{equation}

The errors account for the Poisson noise of the light curves only, and not for the red noise character of the variability process.
The fractional variability amplitude of the SXT, FUV, and NUV light curves of NGC~4593 are     $30.04 \pm 0.82$\%,  $8.30 \pm 0.45$\% and $4.35 \pm 0.21$\%, respectively. Our results show that the source is significantly much more variable in the X--rays than in the UV bands. Within the UV band, the FUV light curves appear to be more variable than the NUV light curve. In fact, the difference between the fractional variability amplitude in the two bands light curves ($F_{var, FUV}- F_{var, NUV}$) is $3.95 \pm 0.49 $\%, which is significant at more than $5 \sigma$ level.
For NGC~7469, the fractional variability amplitude is $11.69 \pm 0.38$\%, $1.20 \pm 0.51 $\% and $0.59 \pm 0.12 $\% for the SXT, FUV and NUV light curves, respectively.
We also calculated the fractional variability of SXT light curves after subtracting the steady sky background count rate (see section \ref{2.2}). This resulted in only a slight improvement in the $F_{var}$ which is not very significant.

As for NGC~7469, the X--ray light curves are more variable than the UV band light curves.
This difference could be due to the fact that the X--ray light curve is longer since the variability amplitude increases with increasing time-scale in AGN. For that reason, we re-calculated $F_{var}$ for the X-ray band, using a 250 ks long light curve (similar to the light curves in the UV bands). We found that the  fractional variability amplitude is  $11.71 \pm 0.43 $\% in this case, which is almost similar to the previous measurement and still significantly larger than the amplitude in the UV bands.
As with NGC~7469, the variability amplitude of the FUV band is larger than the amplitude in the NUV. However, the difference between the fractional variability amplitude in these bands is  $0.61 \pm 0.52 $\%, which is not statistically significant.

\begin{figure*}[!ht]
\includegraphics[width=18cm,height=7.5cm]{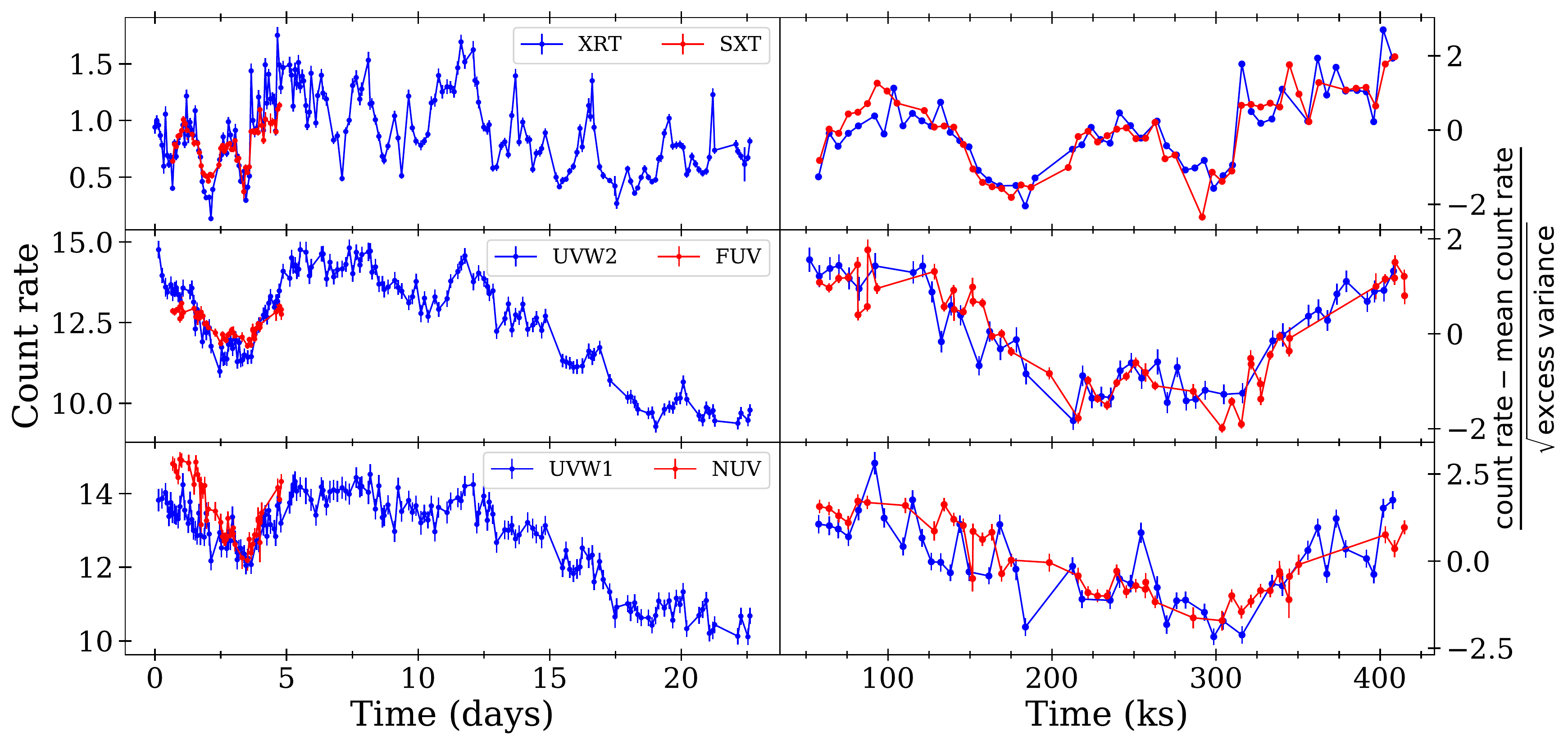}
\caption{Comparison of \astrosat{} and \swift{} X-ray/UV light curves of NGC~4593. \textit{Left panel}: The $\sim 4$ days long \astrosat{} light curves  (SXT, FUV, NUV; in red, from top to bottom panels, respectively) and the $\sim 22 $ days long \swift{} light curves (XRT, UVW2, UVW1; in blue, from top to bottom, respectively). \textit{Right panel}: The same light curves as in the left panel, plotted over the period when both \astrosat{} and \swift{} was observing the source. The mean count rates have been subtracted from the light curves, and the light curves are normalized to the square-root of the excess variances (see text for details).}
\label{swift_astrosat_lc}
\end{figure*}


\begin{figure*}[!ht]
\centering
\begin{subfigure}{0.48\textwidth}
\includegraphics[width=1\linewidth]{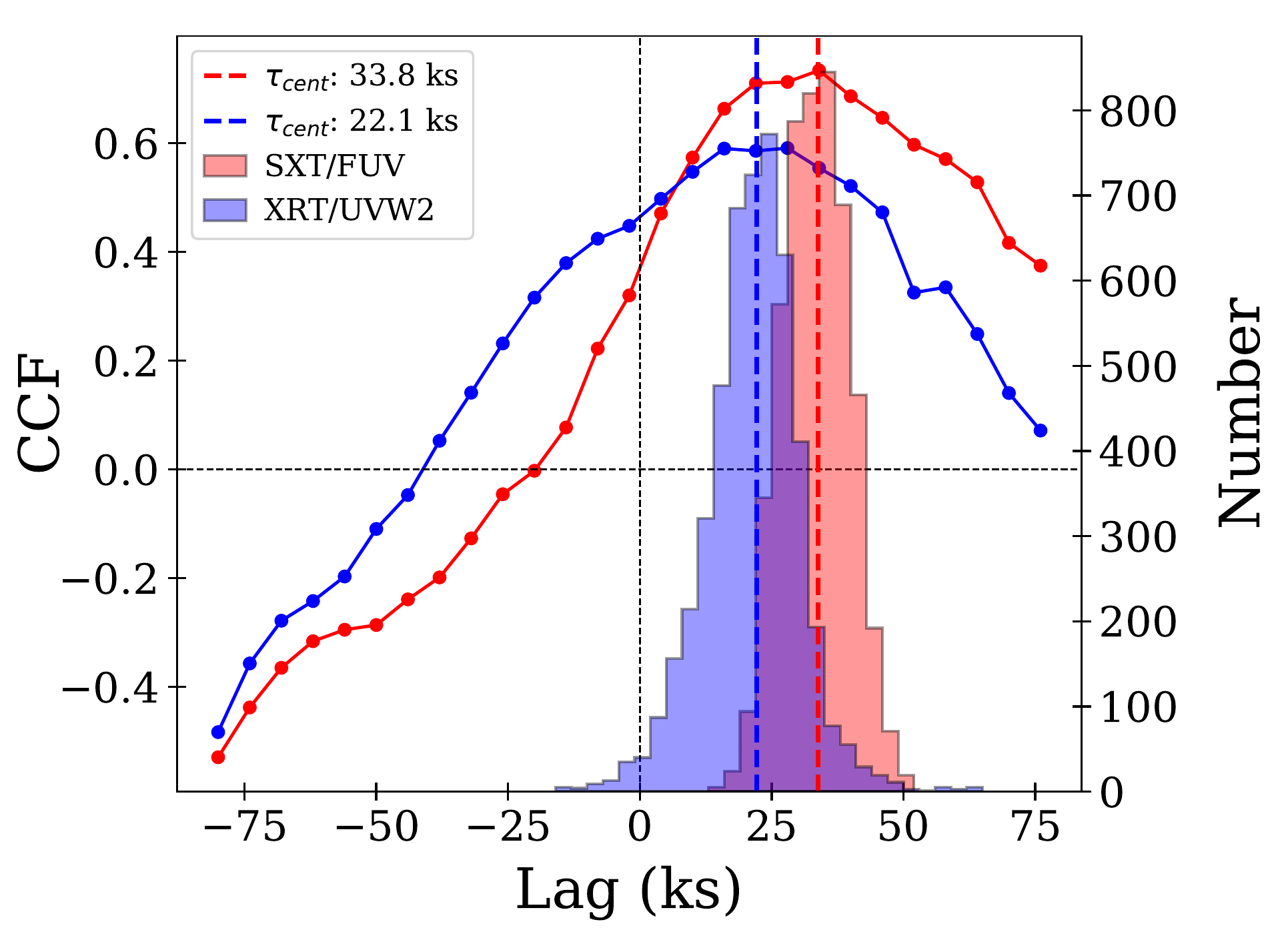}
\caption{}
\label{iccf_sxtfuv}
\end{subfigure}
\hfill
\begin{subfigure}{0.48\textwidth}
\includegraphics[width=1\linewidth]{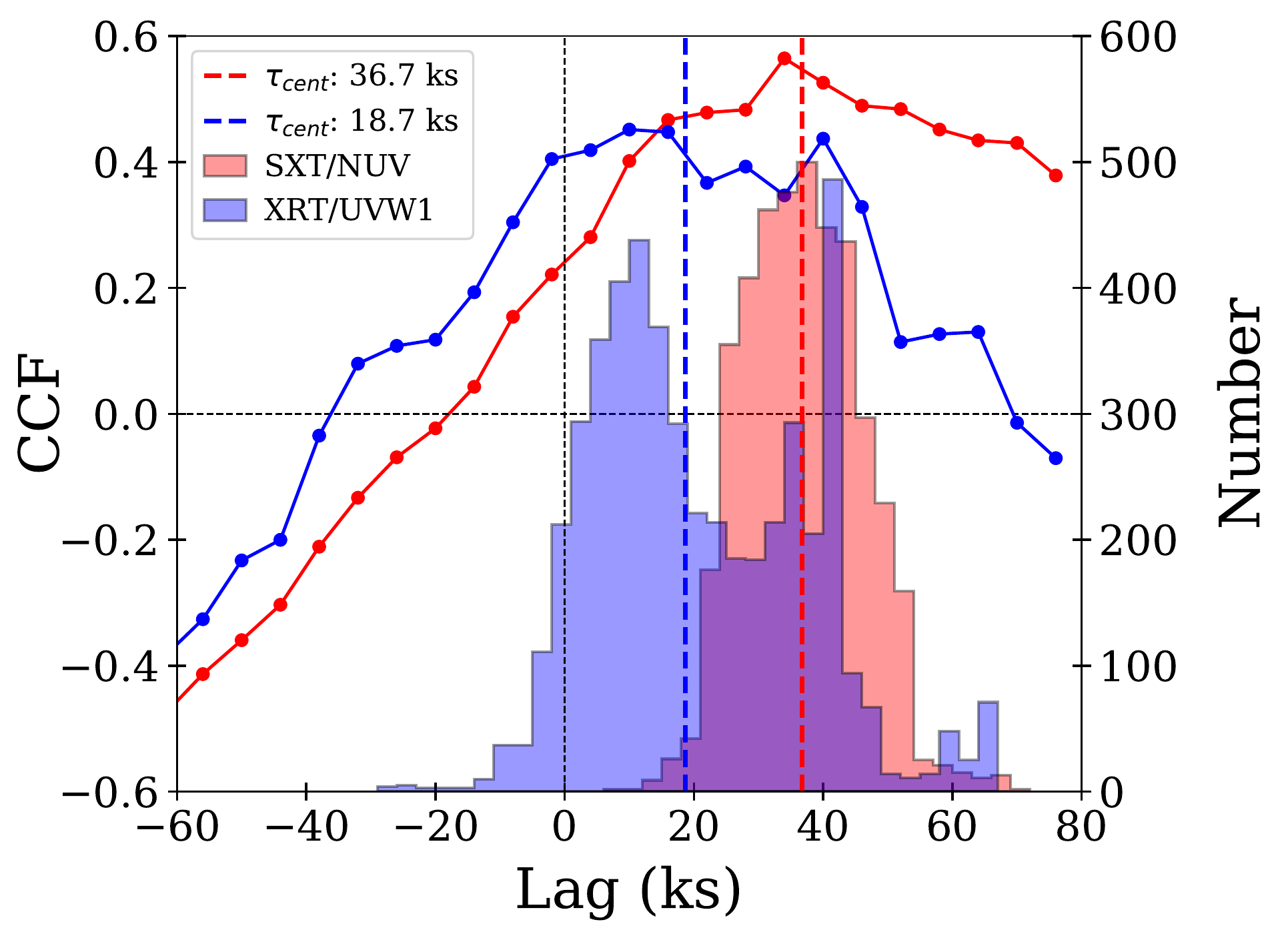}
\caption{}
\label{iccf_sxtnuv}
\end{subfigure}
\caption{The ICCF correlation functions and the centroid lag distributions for the X--ray/FUV and X--ray/NUV light curves (left and right panels, respectively), for the contemporaneous \swift{} data (in blue) and \astrosat{} data (in red) of NGC~4593.  The vertical dashed lines indicate the mean centroid lag, which we accept as our estimate of the delay between the various bands.}

\label{iccf_lag}
\end{figure*}

Lastly, we notice that the variability amplitude of the NGC~4593 X--ray light curve is larger than the amplitude of the NGC 7469 light curve, although their duration is comparable. It is difficult to reach conclusions when we compare $F_{var}$ values which have been measured using single light curves, due to the statistical properties of the distribution of this statistic (see, e.g. \citealt{allevato2013measuring}). Nevertheless, this difference could indicate that the mass of the central black hole in NGC~4593 is smaller than the black hole mass in NGC~7469, as the variability amplitude does depend on the black hole mass as well (e.g. \citealt{papadakis2004scaling, zhou2010calibrating, ponti2012caixa}).

\section{Cross-correlation analysis of NGC~4593}\label{section_ngc4593}

NGC~4593 was observed by \swift{} for nearly
every orbit (96 min) from 13 to 18 July 2016 (6.4 days) and every second orbit for another 16.2 days. These data have already been studied in detail by \cite{mchardy2018x} which can be referred for the observational details. The \swift{} light curves were taken from  \cite{mchardy2018x}. We compared the \swift{} and the \astrosat{} data by over-plotting the XRT (0.5--10.0 keV) with SXT (0.5--7.0 keV), UVW2 ($\lambda_{mean}=$ 1928 \AA) with FUV/BaF2 ($\lambda_{mean}=$ 1541 \AA) and UVW1 ($\lambda_{mean}=$ 2600 \AA) with NUV/NUVB4 ($\lambda_{mean}=$ 2632 \AA) light curves in Fig.~\ref{swift_astrosat_lc} (top, middle and bottom panels, respectively). We shifted the \astrosat{} light curves vertically so that they match with the respective \swift{} light curves in the left panels of Fig.~\ref{swift_astrosat_lc}. These panels show that the \astrosat{} data match quite well with the \swift{} light curves. 

We further checked the agreement between the two sets of light curves by plotting the same light curves in the simultaneous observation window on the right-hand panel of Fig.~\ref{swift_astrosat_lc}.
We subtracted the mean and normalized each light curve with the square-root of the excess variance. The reason for this normalization is that, when computing the Discrete Cross-Correlation Functions (DCF; \citealt{edelson1988discrete}), we subtract the mean and divide the points with the square-root of the excess variance. The right panel in Fig.~\ref{swift_astrosat_lc} clearly shows that the \swift{} and \astrosat{} light curves are in agreement with similar variations, with almost identical amplitude, at all sampled times. We used the light curves plotted in the right  panels (Fig.~\ref{swift_astrosat_lc}) to compute the cross-correlation function between the X--ray and the FUV, NUV light curves.

The measurement errors on the count rates are real, however, as the exposure time in each time bin may not be uniform, some of the abruptly large errors corresponding to smaller exposure times can adversely affect the CCF values. So, we have used the weighted errors for our CCF analysis (for details, see section \ref{section ngc7469}).

\begin{table*}[!ht]
\centering
\renewcommand{\arraystretch}{2}
\caption{Cross-correelation results using the \astrosat{} light curves of NGC~4593 (errors correspond to 68\% confidence region for each parameter).}
\begin{tabular}{P{0.08\linewidth}P{0.11\linewidth}P{0.11\linewidth}P{0.15\linewidth}P{0.11\linewidth}P{0.11\linewidth}P{0.11\linewidth} } 
\hline
\hline
\textbf{Method} &\multicolumn{3}{c}{\textbf{Mean  $\mathbf{\tau_{cent}}$ (ks)}} &\multicolumn{3}{c}{\textbf{Mean CCF$\mathbf{_{cent}}$}}     \\
 & \textbf{SXT/FUV} & \textbf{SXT/NUV} & \textbf{FUV/NUV } & \textbf{SXT/FUV} & \textbf{SXT/NUV} & \textbf{FUV/NUV }    \\
\hline
ICCF & $33.8^{+6.1}_{-6.1}$ & $36.7^{+9.0}_{-9.0}$
& $9.5^{+6.3}_{-8.2}$  & - & - & -  \\
DCF & $43.3^{+5.0}_{-10.0}$ & $52.1^{+10.0}_{-17.5}$ &  $14.8^{+4.0}_{-12.5}$& $0.73^{+0.06}_{-0.07}$ & $0.65^{+0.09}_{-0.10}$  & $0.86^{+0.01}_{-0.05}$  \\ 
\hline
\hline
\end{tabular}
\label{table_ngc4593}
\end{table*}

\begin{figure}[!ht]
\includegraphics[width=\columnwidth]{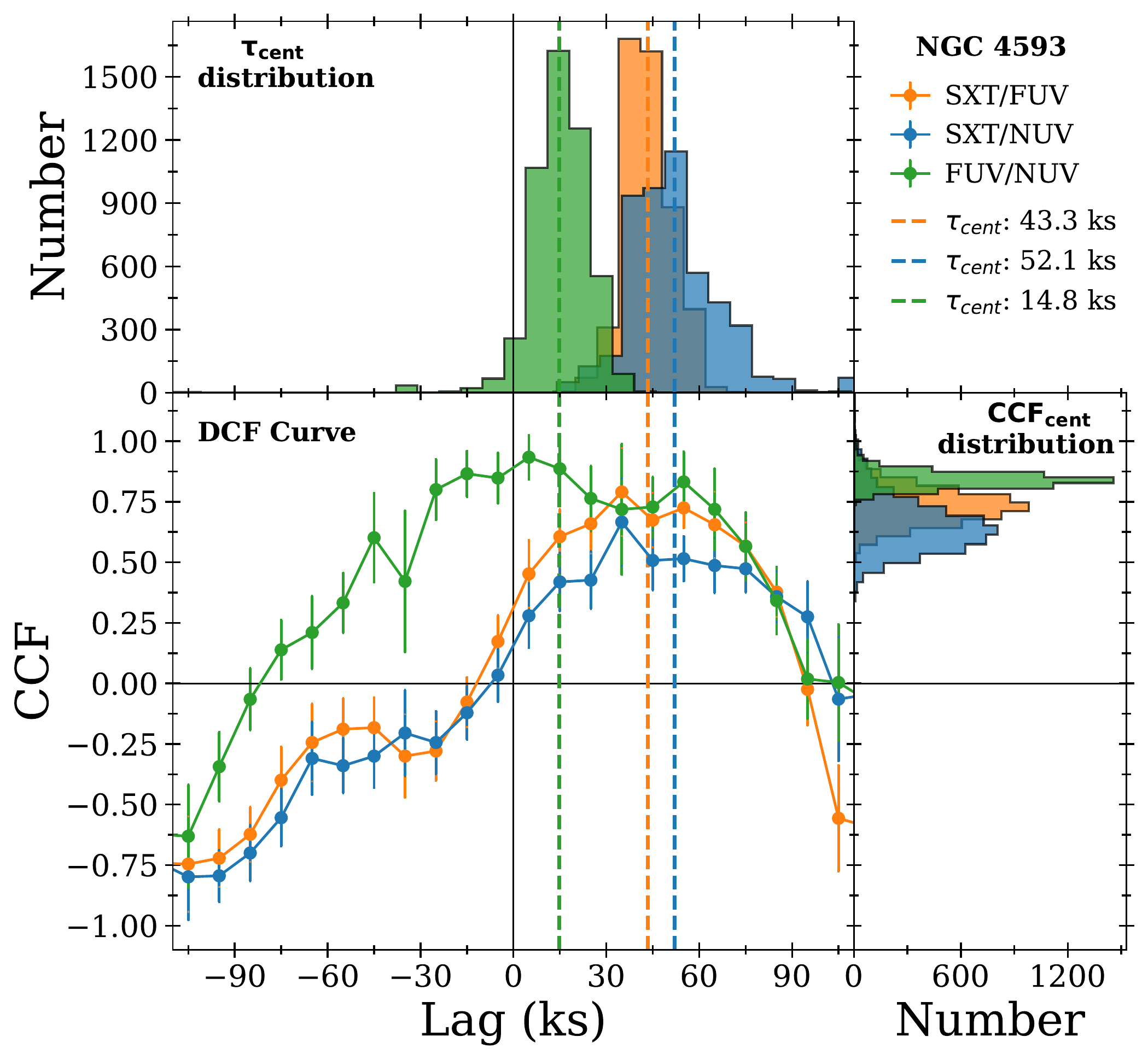}
\caption{The Discrete cross--correlation functions for NGC 4593, using the \astrosat\, light curves, \textit{Lower left panel:} DCF curves, \textit{Lower right panel:} centroid CCF distributions and \textit{Upper left panel:} centroid lag distributions for 5000 \textit{bootstrap} realizations. Distributions in orange, blue and green correspond to SXT/FUV, SXT/NUV and FUV/NUV cross-correlation outputs, respectively. The vertical dashed line indicates the centroid lag ($\tau_{cent}$) for the observed light curves.}
\label{DCF_ngc4593}
\end{figure}

First, we used the Interpolated Cross-Correlation Function (ICCF) technique to compute the CCF between the X-ray and UV light curves (throughout the paper, positive delays mention a delay of the lower energy band with respect to the higher energy light curve). This technique does not use measurement errors in the CCF calculation. We used the Python based package called PyCCF \citep{sun2018pyccf, peterson1998uncertainties}. This code generates the CCF, the centroid time--lag, $\tau_{cent}$, which is the mean of the time--lags for the CCF points larger than $>0.8$ CCF$_{max}$, as well as the $\tau_{cent}$ distribution using the Monte Carlo simulations.
One can  choose for either flux randomization (FR), random subset selection (RSS) or FR/RSS sampling of the light curve. We performed 5000 Monte Carlo realizations of flux randomization (FR) and random subset selection (RSS) to generate the $\tau_{cent}$ distribution and estimate the mean $\tau_{cent}$ and its uncertainty. We accept the mean $\tau_{cent}$ as the best estimate of the delay between the observed variations in various bands.
Fig.~\ref{iccf_lag} shows the resulting correlation functions (red circles) and the corresponding time--lag distributions (red histograms) for the X-ray/FUV and the X-ray/NUV correlations. Our results are listed in Table~\ref{table_ngc4593}.  Errors listed in this table (and throughout the paper) indicate the 68\% confidence limits for each parameter. Our results imply that X--rays lead the UV bands, while the FUV and NUV variations appear to occur almost simultaneously.

In Fig.~\ref{iccf_lag}, we show the ICCF curves and the corresponding distributions of the centroid time--lags in blue color which we derived using the contemporaneous \swift\, light curves (see the blue curves in the right panel of Fig.~\ref{swift_astrosat_lc}).
Our results in this case are: $\tau_{XRT/UVW2} = 22.1_{-8.8}^{+8.4} $ ks,  $\tau_{XRT/UVW1} = 18.7_{-14.5}^{+21.3} $ ks  and $\tau_{UVW2/UVW1}  = 5.0_{-10.9}^{+9.1} $ ks. They are entirely consistent (within the error) with the results we obtained using the \astrosat\, light curves. This is not surprising, given that \swift\, and \astrosat\, light curves are very similar (right panels in Fig.~\ref{swift_astrosat_lc}).
We also noticed hints of a secondary peak in the XRT/UVW1 ICCF curve  and a bimodal distribution in the corresponding centroid time--lag distribution, similar to that found by \cite{beard2022timescale} using the full length ($\sim 22$~days) of \swift{} light curves. However, we did not see such a bimodal distribution in the case of SXT/NUV as well as SXT/FUV and XRT/UVW2  lags distributions, using the same CCF estimation technique. Since the energy bands of the \astrosat{}'s SXT/NUV and \swift{}'s XRT/UVW1 are very similar, and the correlation was computed using contemporaneous light curves, the presence of the bimodal lag distributions in the \swift{} results could be due to some particular trends in the UVW1 (or the XRT) light curve, which, nevertheless,  are not clear to spot.

We also computed the time--lags for the \astrosat{}'s X-ray and UV light curves using the DCF \citep{edelson1988discrete} technique, for which we have used PyDCF\footnote{\url{https://github.com/astronomerdamo/pydcf}} for our analysis. 
We have computed the DCF function between the -120 ks to 120 ks range and kept the bin width equal to $\sim10$~ks. The lag range and bin width have been chosen after different trials with different input values. We have used the default (\textit{`slot'}) weighting scheme in the available DCF code which gives equal weight to all the data points in the lag bin. Here we take into account the measurement errors as well (see eq. 3 in \citealt{edelson1988discrete}). We quote the centroid time--lag which corresponds to the centroid DCF (DCF$_{cent}$) value instead of the peak DCF value, DCF$_{max}$. DCF$_{cent}$ is calculated by averaging over the DCF values $> 0.7$ DCF$_{max}$. To obtain the lag distribution and uncertainties in the lag we have used 5000 pairs of $\textit{bootstrap}$ realizations using random subset selection.

We plot our DCF correlations in Fig.~\ref{DCF_ngc4593} (left and right panels, respectively), and we list our results in Table~\ref{table_ngc4593}. Orange, blue and green colours indicate the results for the SXT/FUV, SXT/NUV and FUV/NUV correlations, respectively. The top and bottom right panels show the distribution of the centroid time--lags and of the centroid correlation coefficients, respectively. The vertical lines in the top and left bottom panels indicate the mean centroid time--lag.

The results listed in Table~\ref{table_ngc4593} show that the cross-correlation results derived from both techniques are fully consistent with each other, within the errors. The strength of the correlation between the FUV and NUV bands is quite strong (the peak CCF is $\sim 0.86$), and it is larger than the strength of the correlation between the X-rays and the UV bands, which actually decreases from the X--ray/FUV to the X--ray/NUV bands. The mean time--lag increases from $\sim 38$~ks ($\sim$ 0.43~days)  to $\sim 44$~ks ($\sim$ 0.51~days), from the X--ray/FUV to the X-ray/NUV correlations. However, due to the large errors, this difference is not statistically significant. Similarly, the FUV/NUV time--lag is positive, approximately $\sim 10$ ks, which is consistent with the difference between the SXT/FUV and SXT/NUV time--lags but again, this is not significantly different than zero. 

The time--lags we estimated using the contemporaneous  \swift\, or  \astrosat{} light curves are shorter than the time--lags we get when we consider the longer \swift\, light curves (shown with the blue points in the left panels of Fig.~\ref{swift_astrosat_lc}). Indeed, using PyCCF and the $\sim 22$ days long \swift{} light curves, we found the mean centroid lag for XRT/UVW2 and XRT/UVW1 to be $0.66_{-0.15}^{+0.17}$ days and  $0.72^{+0.24}_{-0.46}$ days, respectively. These lag estimates are consistent with the results of \cite{mchardy2018x} and \cite{cackett2018accretion}. The time--lag estimates from the shorter light curves are $\sim1.5$ (within uncertainty) times smaller than the lag values for the full light curves. This can be due to the short observations that can not capture the longer timescale variations \citep{beard2022timescale}, or even the  physical reasons that we discuss in \S~\ref{discussion}.


\section{Cross-correlation analysis of NGC~7469}\label{section ngc7469}

\begin{figure}[!ht]
\includegraphics[width=\columnwidth]{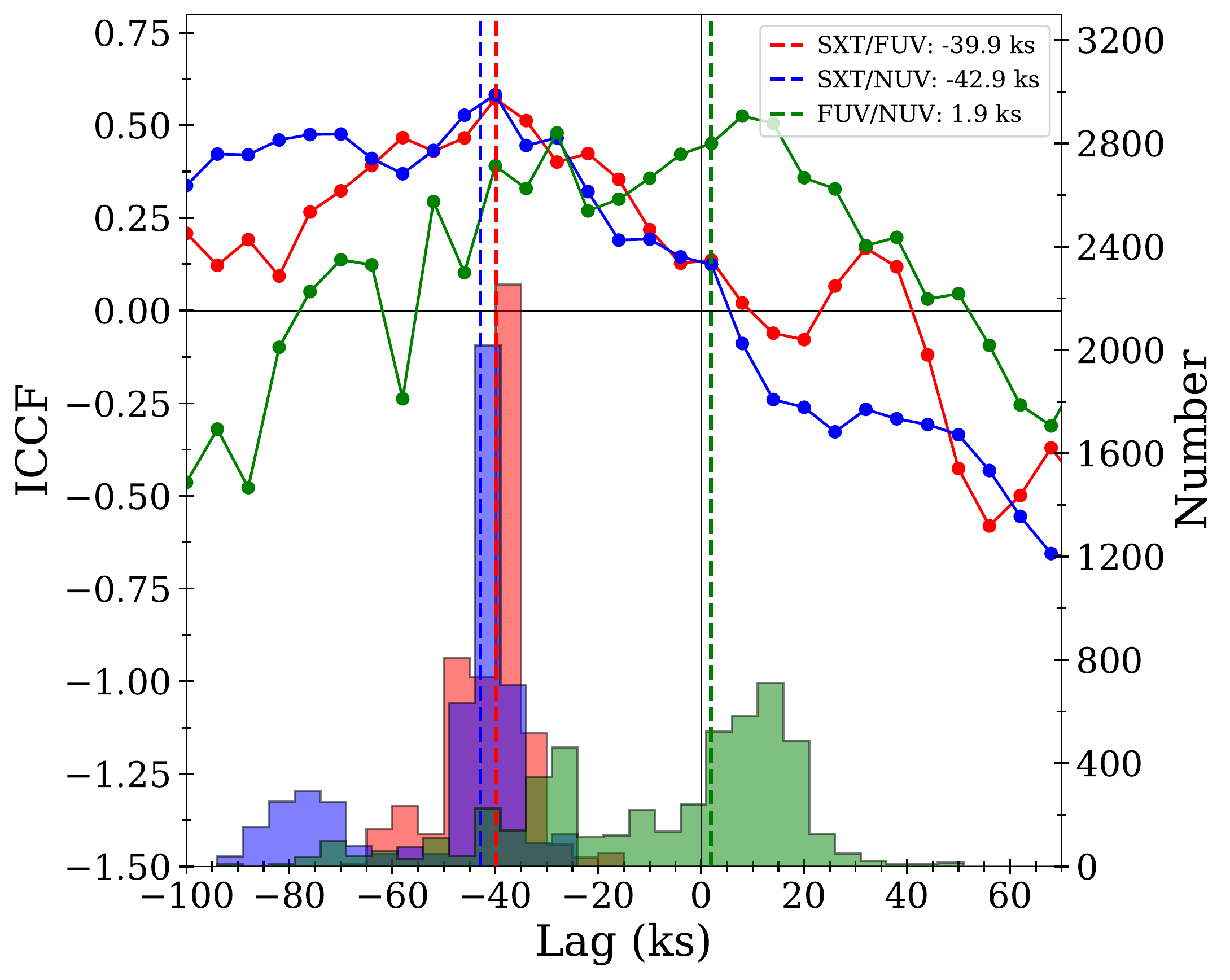}
\caption{Cross-correlation plots for NGC 7469, between the SXT/FUV, SXT/NUV and FUV/NUV bands (red, blue and green colours, respectively), using the ICCF technique (PyCCF). 
Vertical dashed lines indicate the centroid lag for the correlation in the various bands.}
\label{iccf_ngc7469}
\end{figure}

We have used the $\sim4$ days long \astrosat{}'s X-ray/UV light curves of NGC~7469, shown in Fig.~\ref{LC_ngc7469}, to compute the delays between various energy bands, using the same techniques as discussed for NGC~4593 in the previous section. 
Fig.~\ref{iccf_ngc7469} shows the cross-correlation results obtained using the ICCF method. 
We notice that the peak for the SXT/FUV and SXT/NUV correlations is centred near $-40$ ks and the other, a weaker peak is located at  approximately $-80$ ks. The important result though is that the lag distributions are clearly shifted to negative time--lags. It strongly indicates that the variability in the X-ray band is lagging behind the variability in the UV bands. 
The centroid lag distribution for FUV/NUV is scattered over a broad lag range but is consistent with zero lag within uncertainty. The mean time--lags for the correlation between the various bands are listed in Table~\ref{table_ngc7469}. 

As before, to verify our results, we further used the DCF technique to obtain the lag and the correlation strength in this case.
We computed the DCF correlation between the -100 ks to 100 ks range and kept the bin width equal to 10 ks in each case.  To find the lag distribution and to calculate the uncertainties on the mean centroid time--lag we have used 5000 pairs of $\textit{bootstrap}$ realizations.

\begin{figure}[!ht]
\includegraphics[width=\columnwidth]{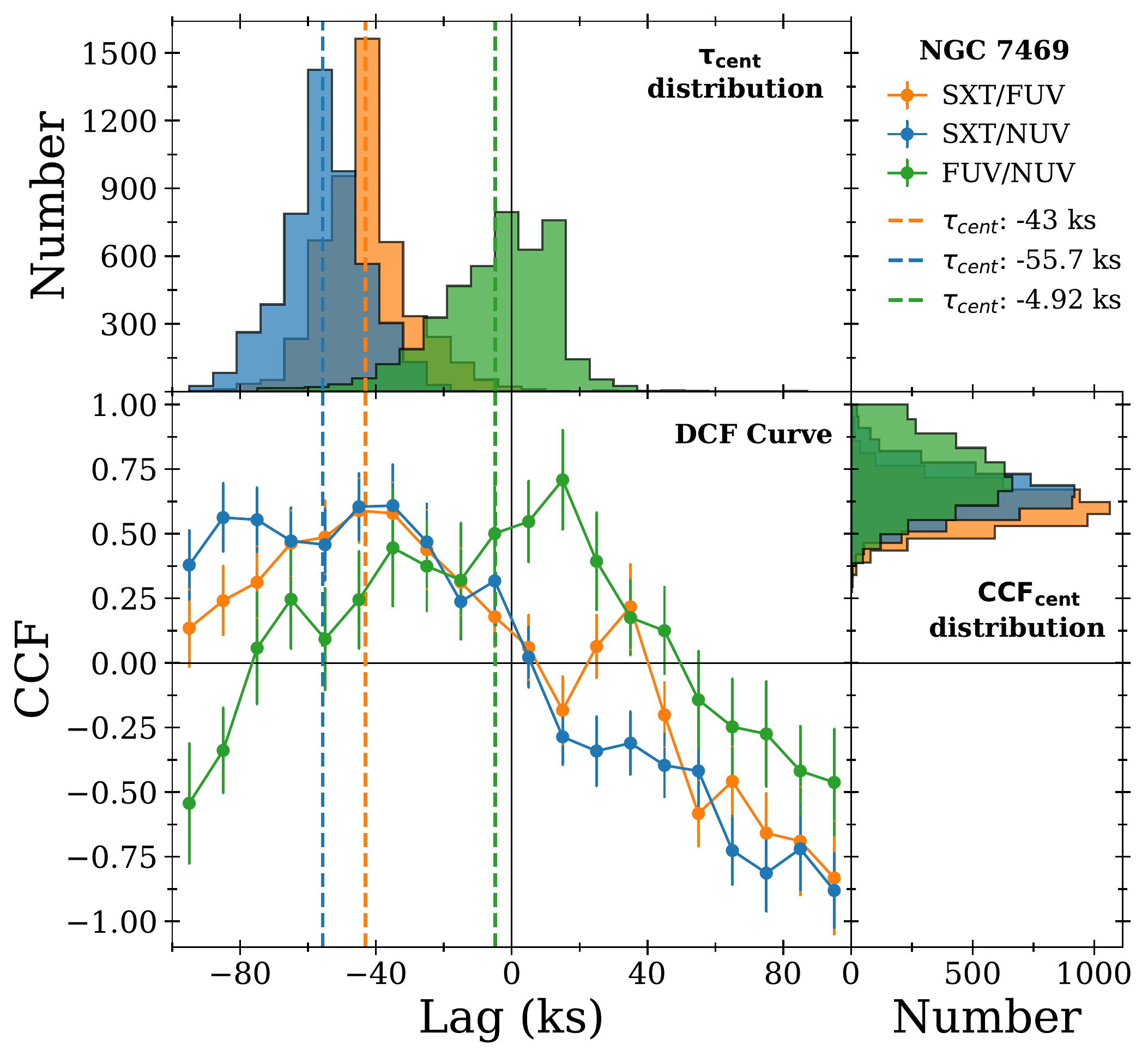}
\caption{Same as Fig.~\ref{DCF_ngc4593} for NGC 7469.}
\label{DCF_ngc7469}
\end{figure}

\begin{table*}[!ht]
\centering
\renewcommand{\arraystretch}{2}
\caption{Same as Table~\ref{table_ngc4593} for NGC~7469.}
\begin{tabular}{P{0.08\linewidth}P{0.11\linewidth}P{0.11\linewidth}P{0.15\linewidth}P{0.11\linewidth}P{0.11\linewidth}P{0.11\linewidth} } 
\hline
\hline
\textbf{Method} &\multicolumn{3}{c}{\textbf{Mean $\mathbf{\tau_{cent}}$ (ks)}} &\multicolumn{3}{c}{\textbf{Mean CCF$\mathbf{_{cent}}$}}     \\

 & \textbf{SXT/FUV} & \textbf{SXT/NUV} & \textbf{FUV/NUV } & \textbf{SXT/FUV} & \textbf{SXT/NUV} & \textbf{FUV/NUV }    \\
\hline
ICCF & $-39.9^{+3.1}_{-8.7}$ & $-42.9^{+5.6}_{-30.4}$ & $1.9^{+12.1}_{-35.9}$  & - & - & -  \\

DCF & $-43.0^{+13.3}_{-10.0}$ & $-55.7^{+11.4}_{-12.0}$ &  $-4.92^{+6.7}_{-26.0}$ & $0.60^{+0.08}_{-0.08}$ & $0.66^{+0.09}_{-0.09}$  & $0.72^{+0.13}_{-0.14}$\\

\hline
\hline
\end{tabular}
\label{table_ngc7469}
\end{table*}


\begin{figure}
\includegraphics[width=\columnwidth]{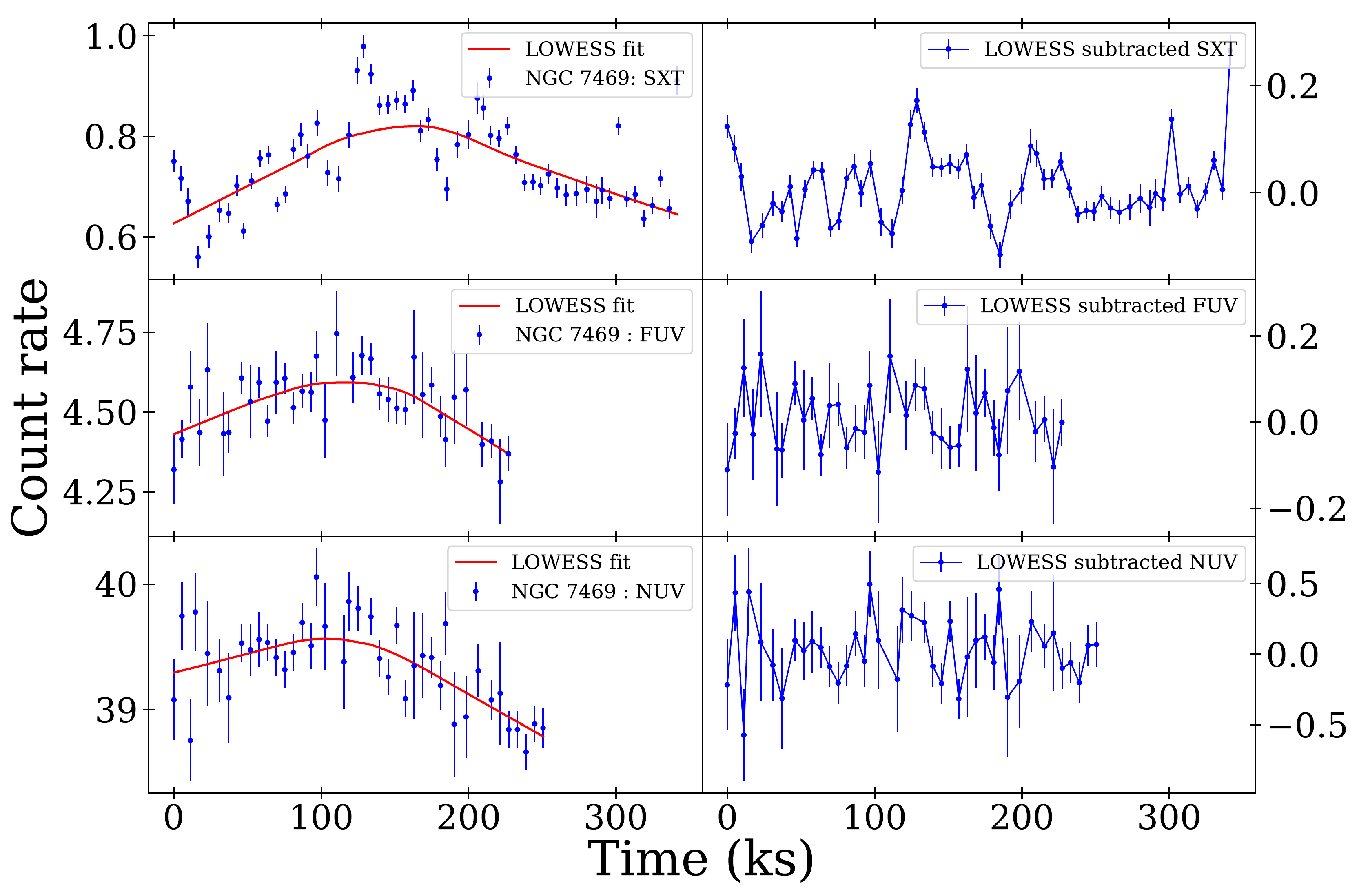}
\caption{ \textit{Left panel:} Light curves of NGC~7469 fitted with LOWESS function (in red). \textit{Right panel:} LOWESS subtracted light curves. }
\label{lowess_ngc7469}
\end{figure}

As mentioned earlier, we compute the DCF using eq. (3) in \cite{edelson1988discrete}, i.e. taking into account, and correcting the light curve variance from the contribution of the Poisson noise. Initially, we found that the DCF peak was $> 1$ for the cross-correlation between FUV and NUV light curves. This is probably due to the fact that  the error of a few data points in the UV light curves is large, due to the short exposure in some orbits (see Fig.~\ref{LC_ngc7469}). When we plot the errors versus the exposure time in each orbit, we find that the errors are inversely proportional to the exposure time. As it is customary, we use the mean square error of all the points as an estimate of the Poisson noise variance. However, although the errors may be correct,  maybe this way of computing the variance due to Poisson noise may not be appropriate in this case, because the contribution of the large errors to the overall variance may be overestimated. For that reason we decided to consider the weighted mean of the experimental errors as a better estimator of the Poisson noise  variance, as follows: $\sigma_{p}^2=(1/N)\sum_{i}^{N}w_i \sigma^2_{err, i}$, where the weight for each point is defined as $w_i=T_{exp, i}/T_{exp,max}$, where $T_{exp,i}$ is the exposure time of each measurement in the light curves. When using this formula for the estimation of the Poisson noise variance we found a reasonable DCF peak (i.e. $DCF_{cent}< 1$) so, we performed the further cross-correlation analysis assuming  the mean of the weighted errors as the appropriate estimator for the Poisson noise variance of each light curve.
With the weighted errors, we found reasonable DCF peak values for FUV/NUV correlation, however, for some ($10-15$\%) of the simulated pairs of FUV and NUV light curves, we still found $DCF_{max} > 1$ which we have discarded from our analysis.
The resulting correlations (together with time--lags and CCF strength distributions)  are plotted in Fig.~\ref{DCF_ngc7469}, and the results are listed in Table~\ref{table_ngc7469}.

The correlation functions as calculated from both techniques are all in agreement. The mean time--lags are $\sim$~41~ks ($\sim$~0.47~days)  and $\sim$~49~ks ($\sim$~0.56~days) for the X--ray/FUV and X-ray/NUV correlations, respectively.
The main result from the CCF analysis of the \astrosat\, light curves of NGC 7469 is that the X--ray variations are {\it lagging} those detected in the UV band. This lag behaviour is consistent with previous results regarding the cross--correlation between the UV and the X--ray light curves of the same source. For example, \cite{pahari2020evidence}
used a month-long X-ray light curve from RXTE/PCA and 1.5 month-long UV continuum light curves from \iue{} spectra and found that  X-rays are delayed relative to the UV continuum by $\sim 3.5$ days. However, when they removed variations slower than 5 days from the X-ray light curves, then they found the UV variations lagging behind the X-ray variations by 0.37 $\pm$  0.14 days days, which is what is usually detected in other Seyferts. 

Although our light curves are only 4 days long, i.e. shorter than the longer time-scales that \cite{pahari2020evidence} removed from their light curves, 
we decided to investigate the effects of the longest time--scales in the light curves. For this reason, we filtered out the variations slower than $\sim$ 2.5 days, and then repeated the cross-correlation analysis.  For filtering or de-trending the light curves, we used the Python based locally weighted scatter plot smoothing function, LOWESS \citep{pahari2020evidence}. 
The red lines in Fig.~\ref{lowess_ngc7469} are the smoothed light curve obtained after applying the LOWESS filter to eliminate the variations slower than 2.5 days.
The red lines  clearly indicate why the cross-correlation analysis suggests a UV lead in this source, for the given data set. 
The resulting FUV and NUV light curves turn out to be non-variable. We again fitted a constant to the filtered light curve and found $p_{nul}=$ 0.73 and 0.32 for the FUV and NUV light curves, respectively. Given this result, we cannot search for correlated variations on short time-scales between X--ray and the UV light curves. If there are variations in the UV bands which are delayed with respect to the X-rays, their amplitude is smaller than the errors of the UV light curves.
However, we also noticed a weaker secondary peak at a lag of $\sim +\rm 35$ ks  in the ICCF and DCF of SXT/FUV light curves (see the red and orange CCF curves in Fig.~\ref{iccf_ngc7469} and \ref{DCF_ngc7469}).  So, although detrending is not possible in our case (for the reasons discussed above), the observed cross-correlation functions do show a hint of X-ray reprocessing time lag similar to that found by \cite{pahari2020evidence} in this source.

\section{Results and discussions}
\label{discussion}

We study the relationship between the variations in the UV and X-ray emission from two Seyfert 1 galaxies namely NGC~4593 and NGC~7469 using 
$\sim 4{\rm~days}$ long X--ray, FUV and NUV light curves, obtained from AstroSat observations.
We used two different techniques -- ICCF and DCF for our cross-correlation analysis and measured time--lags between X-ray and UV bands. The main results are as follows.

\begin{enumerate}[(a)]
\item  The \astrosat{} light curves of NGC~4593 agree very well with the \swift{} light curves in the common observation window (see Fig.~\ref{swift_astrosat_lc}). The X-ray/UV lags measured using the \astrosat{} and the synchronous \swift{} light curves are also similar  within the uncertainties (see Fig.~\ref{iccf_lag}). This demonstrates \astrosat{}'s capability to study X-ray/UV correlations in AGN.

\item In the case of NGC~4593, variations in the X--ray light curve leads those in the UV light curves by $\sim 0.5$~days (see Table \ref{table_ngc4593}).  

\item In contrast to NGC~4593, we find the opposite lag behaviour in NGC~7469 where the variations in the UV emission lead those in the X-ray emission by $\sim$ 0.5~days (see Table \ref{table_ngc7469}). 

 \item The FUV and NUV band emission from both the sources are strongly correlated without any significant delay.

\end{enumerate}

\subsection{Cross-correlation strength}

The centroid peak of CCF curves between the X-ray and UV light curves is $\sim 0.7$ for NGC~4593 and $\sim 0.65$ for NGC~7469 (see Table \ref{table_ngc7469}). These X-ray/UV correlations are somewhat weaker than the FUV and NUV correlations which show CCF peaks of $\gtrsim 0.8$ (within uncertainty) for both sources. The weaker correlation strength between X-rays and UV has been observed previously by many researchers (see e.g. \citealt{edelson2019first}). Indeed, it has been observed that the X-ray/UV correlation strength is in general  weaker ($\rm{CCF_{peak} < 0.75}$) than UV/optical correlations ($\rm{CCF_{peak} > 0.8}$), however, it is elusive to comment whether this is an intrinsic property of AGN variability or simply an artefact of the CCF analysis \citep{edelson2019first}.

In the X-ray reverberation model, reprocessed optical/UV light curves are the convolution of variable X-rays with the response function of the accretion disk. This naturally leads to smoother optical/UV light curves than X-ray light curves and this should naturally lead to a lower CCF peak between X-rays and the optical/UV bands. 
Recently, \cite{panagiotou2022explaining}  showed that a low X-ray/UV correlation can be expected when the dynamic variability of the X-ray source is taken into account. In particular, these authors showed how the geometric or physical configuration of the X-ray source can affect the expected correlation strength. The variations of the geometric configuration (the X-ray source height, for example) can result in a weak X-ray/UV correlation, while the correlation between the UV and optical variability remains strong.


As for NGC~7469, in which we observe the lead of UV photons, we clearly see that there are short and medium term variations in X-ray light curves that do not appear in the UV light curves (and for that reason $F_{var,xray} >> F_{var,uv}$). This shows that the corona varies not only due to the input UV variations but also due to other factors (associated with the physical processes powering the corona and X-ray emission mechanism). This significant difference between the variability amplitude in X-rays and in the UV bands can certainly explain (to some extent) the fact that we do not see a very strong correlation between the X-ray and UV light curves.

\subsection{The X--ray/UV correlation in NGC~4593}

Using $\sim$ 4 days long \astrosat{} data on NGC~4593, we observed that the variability in the soft X-ray band is leading the FUV and NUV bands by $\sim$ 38 ks and $\sim$ 44 ks, respectively. 
These results are in agreement with the hypothesis of the X--ray illumination of the accretion disk in this object. 

We found that the lag values obtained  for the \astrosat{}, and the contemporaneous \swift{} light curves are in agreement (within uncertainty). However, these lag values are $\sim$ 1.5 times smaller if we compare with the lags  obtained using the full (22 days long) \swift{} light curves \citep{mchardy2018x, cackett2018accretion}. One possibility for the  smaller lag that we find is that, the shorter \astrosat{} light curves do not sample the longer time-scale variations in the light curves. This aspect has been studied in detail in \cite{beard2022timescale}.

However, there may be another explanation for the fact that our delays are slightly shorter than the delays measured when using the longer \swift{} light curves. Fig.~\ref{swift_astrosat_lc} shows that the X-ray flux during the first part of the \swift{} long observation was lower than in the rest of the observation. This could happen if the luminosity of the X-ray source was  smaller than the average luminosity of the X-ray source in the remaining part of the observation. The X--ray/UV time--lags depend on the luminosity of the X-ray source. For example, Fig. 23 in \cite{kammoun2021uv} shows that, as the X-ray luminosity decreases, the X-ray/UV time--lag also decreases, with the difference being more pronounced in the case when the spin is high. This could then explain the slightly smaller delay we measure with the \astrosat{} observations in this source.

\subsection{The X--ray/UV correlation in NGC~7469}\label{sec:6ngc7469}

Although our results are in agreement with the past results, in the sense that, we also detect the UV variations to lead X--rays in this object, thus favouring the up-scattering of UV photons from the hot corona as the dominant process for the production of the X--rays in this source, our time--lag value of $\sim$ 0.5 days is smaller than the lag observed by \cite{nandra1998new} and \cite{pahari2020evidence}.  

Time--lags between UV and X--rays due to thermal Comptonization should be representative of the light travel time between the UV emitting region and the X-ray source, as well as the time it takes for the optical/UV seed photons to be up-scattered by the hot electrons to X-ray energies.  Regarding the first term, we would like to point out that the FUV and NUV photons detected by the observer (and the X--ray corona) do not originate from a single disk annulus. Instead, as \cite{kammoun2021uv} have shown, these photons should originate from a large area in the inner disk, and the transfer function of the X-ray corona response to disk photons instantaneous emission should be identical to the disk transfer functions presented in that paper. The accretion rate ($\dot{m}$) and BH mass of NGC~7469 are $\sim 0.3$ \citep{pahari2020evidence} and $\sim 10^7$ M$_{\odot}$ (\citealt{peterson2014reverberation}), respectively. According to \cite{kammoun2021uv}, we would expect the delay between $\sim 2000$ \AA\, photons and the X--rays to be of the order of $\sim 0.5$ days, if we assume a spin of zero, just from light travel effects (see the right panel in their Fig.~17). This is entirely consistent with our results.

As we already mentioned, \cite{pahari2020evidence}, \cite{vincentelli2021multiwavelength} have observed that the UV/X--ray lead can change after filtering slower variations (> few days) from the light curves. \cite{pahari2020evidence} have found that after filtering slower variations (> 5 days) from the X-ray light curve of NGC~7469, the UV (1315 \AA) variability lags the X-ray variability by 0.37 $\pm$  0.14 days, which is consistent with the disk reprocessing. We have already shown that such filtering or de-trending is not possible for our \astrosat{} light curves, due to the short  observation period ($\sim 4$ days) and the low variability amplitude of the UV light curves, but, we do notice a hint of positive lag in the SXT/FUV cross-correlation functions.

However, the non-detection of the X--ray disk thermal reverberation lags (without filtering slower variations) in NGC~7469 is puzzling, as they are detected in almost all AGN with long, well sampled light curves (including NGC 4593, as presented in this work, and many other works in the past).

One possible explanation can be given if we look at Fig.~25 in \cite{kammoun2021uv}. We notice that the reverberation fraction ($R_{rev}$) can be as low as $\sim$ 5\% (for $\dot{m} = 0.3$, a=0, $\lambda \sim 2000$~\AA), which means that the variable X--ray thermal reverberation signal in the UV light curves should be very weak in NGC~7469, and that's why we do not detect it with the cross-correlation analysis of the \astrosat{} observations, which are not as long as the past observations. The reason for weak $R_{rev}$ is that for a higher accretion rate ($\dot{m}$) the disk is hotter and hence the disk emission (in a wave band) is larger which leads to a smaller reverberation signal. 
We believe that positive time--lags will be detected in future \astrosat{} observations, as long as they are comparable to the observations in the past. 
If true, this result suggests that the observed UV flux variability in NGC~7469 is probably due to the \textit{intrinsic} disk variability.

Finally, we note that, in a way, it seems surprising that UV lead over the X--ray variations are not observed more frequently in the cross-correlation analysis of X-ray/UV/optical light curves of Seyferts, if inverse Compton scattering of disk photons is the physical process which produces X--rays in AGN. Even if the intrinsic disk emission is not variable (or it is variable but with small amplitude, as our analysis shows), X--rays are variable in all AGN. If they illuminate the disk, they should produce a variable component in the disk emission at each wavelength. These variable soft photons will then re-enter the corona to be up-scattered, and hence they should introduce a delay in the X--ray emission, equal (but with a negative sign) to the delay that X--rays imprint to the disk emission at each wavelength. 

However, the majority of the variable disk emission at each $\lambda$ may not be easy to detect
when combined with the X--ray variations induced by all the other disk variable photons entering the corona, and, in particular,  when  diluted by the intrinsic, large amplitude X--ray variations. In other words, when we correlate X--rays with various disk bands, we detect the UV delays because the main drivers of the observed variations in each band are the illuminating X--rays. On the other hand, it is much more difficult to detect the UV leading the observed X--ray variations, because the latter is mainly determined by other factors. 

It may be the case that, in NGC 7469, the FUV and NUV bands are close to the mean energy of the disk photons that enter the corona, they are intrinsically variable, {\it and} the intrinsic X--ray variations are not as large as in the other AGN. Indeed, the X--ray variability amplitude is $\sim 3$ times smaller in NGC 7469 than in NGC 4593 (see \S~\ref{sec:fvar} ). It is this combination of factors that may allow us to detect the UV lead on the X--ray variations in this object.


\section{Conclusions}\label{conclusion}

We analysed nearly four days long \astrosat{} UV/X-ray observations of NGC~4593 and NGC~7469. We found that the UV/X-ray emission from  both the sources are variable, with X-rays being more variable than UV emission (see \S~\ref{sec:fvar}). 
For NGC~4593, we found that the \astrosat{} light curves are in full agreement with the contemporaneous \swift{} light curves (Fig.~\ref{swift_astrosat_lc}). Both missions revealed similar UV lags relative to X-rays that favour the disk reprocessing model as found by earlier researchers. 
As for NGC~7469, we found that the variations in X-ray light curves are lagging UV by $\sim0.5$ days (Fig.~\ref{iccf_ngc7469} and \ref{DCF_ngc7469}). This can be explained as the light travel time between the accretion disk and corona under the assumption of lamp-post geometry. However, the non-detection of the thermal reverberation signal in NGC~7469 is intriguing.  For a source similar to NGC~7469 having a high accretion rate ($\dot{m}=0.3$), the disk 
is expected to be hotter (for low spin), thus giving rise to stronger disk emission (in a wave band)  which leads to a weaker reverberation signal ($R_{rev}$). This could be the reason for the absence of reprocessing signals in  our short \astrosat{} observations, and the observed UV variability is most likely intrinsic to the disk. 
This result makes our observation important because it is generally difficult to detect the \textit{intrinsic} disk emission due to the combined effect of X-ray variations. Future observations with longer exposure time will be required to further explore this source.

\section{Acknowledgement}
This research has made use of the data from the \astrosat{} mission of the Indian Space Research Organisation (ISRO), archived at the Indian Space Science Data Centre (ISSDC). 
This work has used the data from the Soft X-ray Telescope (SXT) developed at TIFR, Mumbai, and the SXT POC at TIFR is thanked for verifying and releasing the data via the ISSDC data archive and providing the necessary software tools.  The UVIT project is a result of collaboration between IIA Bengaluru, IUCAA Pune, TIFR Mumbai, several centres of ISRO and CSA. The UVIT data were checked and verified by the UVIT POC at IIA, Bangalore, India. The UVIT data were processed by the CCDLAB pipeline. This research has made use of data obtained from the 4XMM XMM-Newton serendipitous source catalogue compiled by the XMM-Newton Survey Science Centre consortium.

\section{Data Availability}
The \astrosat{} data are
available at \url{https://astrobrowse.issdc.gov.in/astro_archive/archive/Home.jsp}. The \swift{} data are available at \url{https://www.swift.ac.uk/archive/}.



\bibliographystyle{mnras}
\bibliography{mybib} 




\appendix

\section{Radial profile analysis}
\label{sec:radprof}


\begin{figure}[!ht]
\centering  
\begin{subfigure}{0.9\textwidth}
\includegraphics[width=1\linewidth]{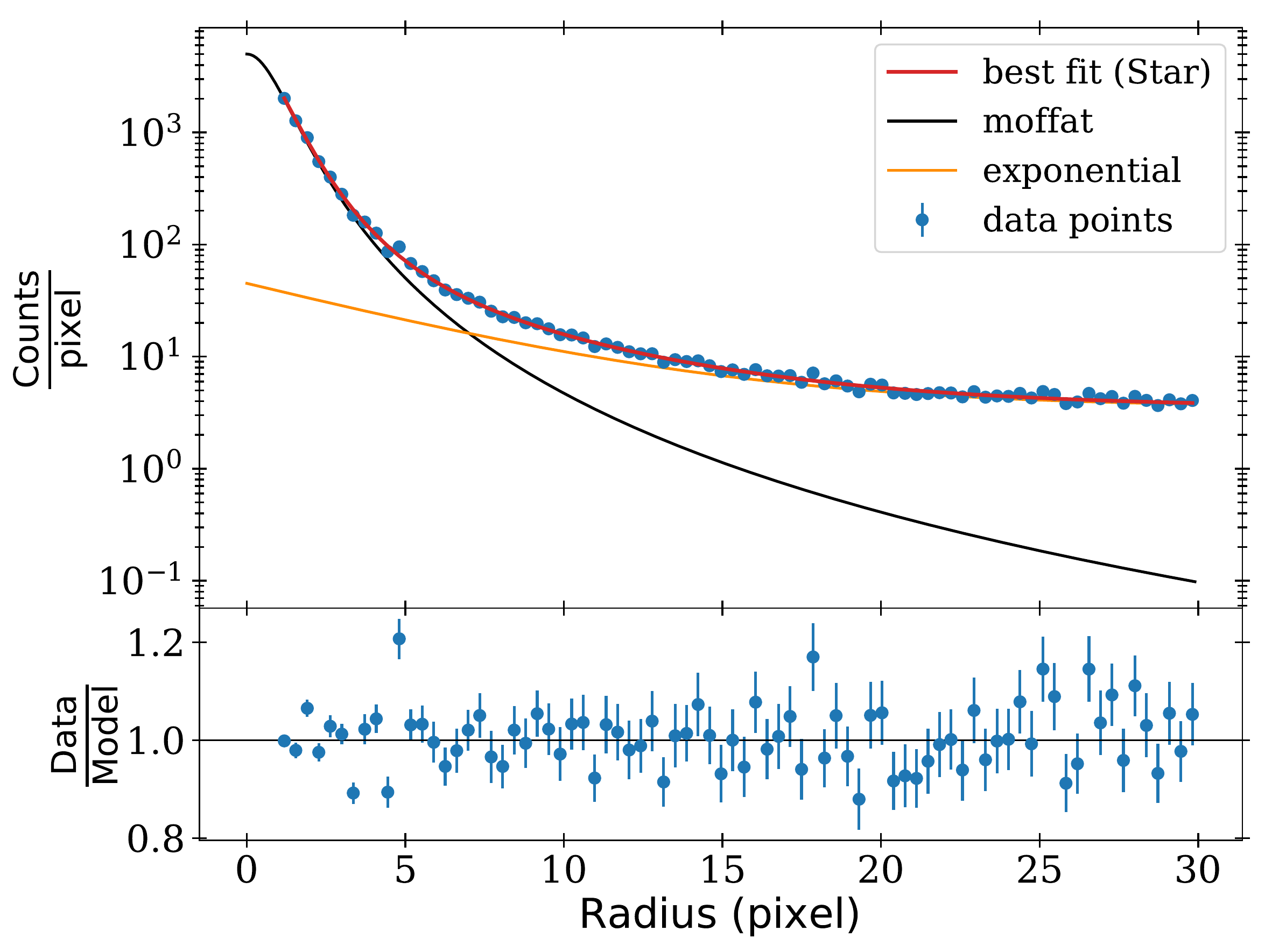}
\caption{}
\label{star_nuv}
\end{subfigure}
\medskip
\begin{subfigure}{0.9\textwidth}
\includegraphics[width=1\linewidth]{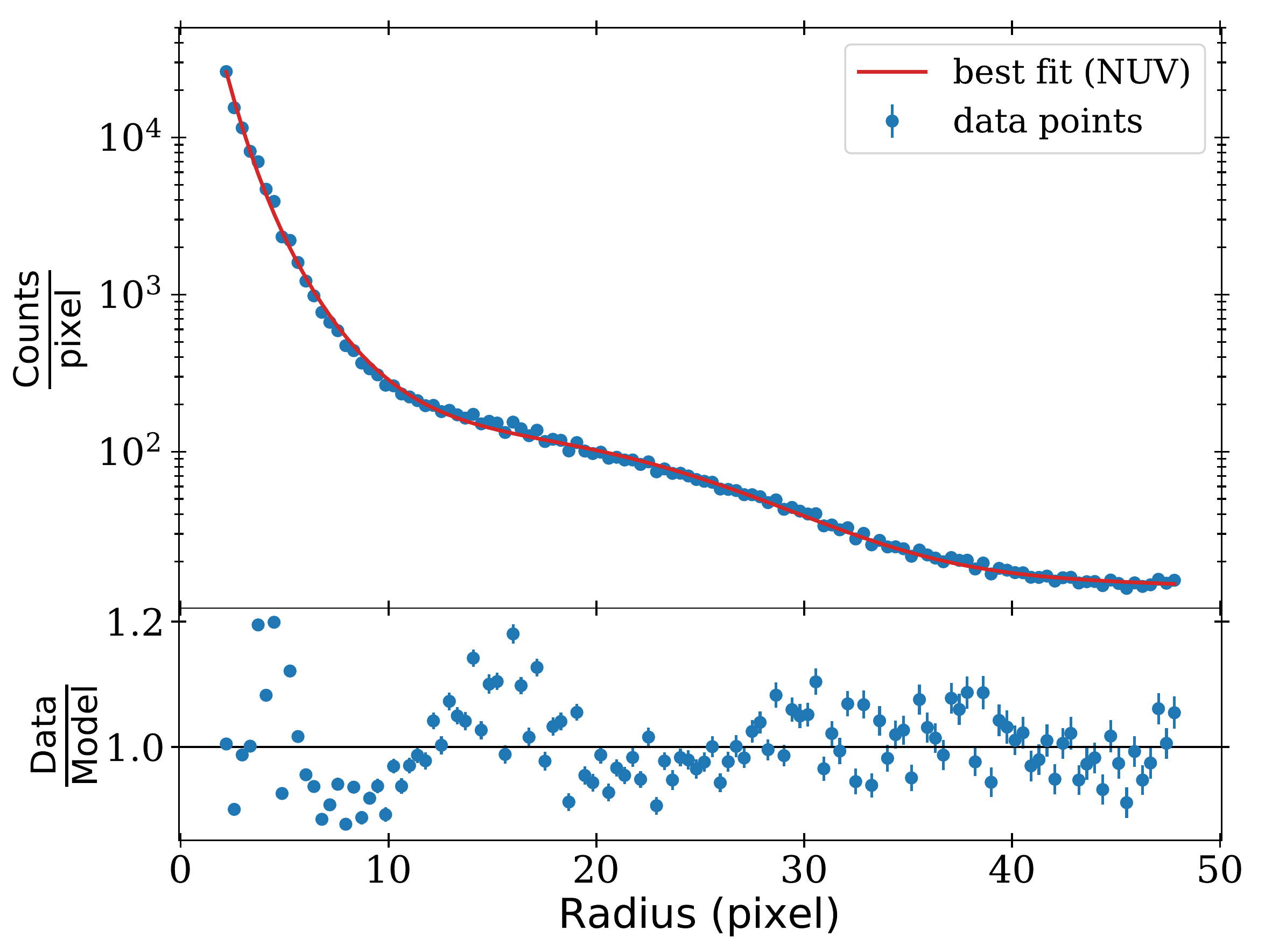}
\caption{}
\label{radial_nuv}
\end{subfigure}
\caption{\small{ Radial profile fit of: (a) bright star (TYC~1160--1473--1) in NUV band, using a Moffat function to fit the core and an exponential function to fit the wings, plus a constant background  (b) NGC~7469 in NUV band. }}
\label{radprofile_nuv}
\end{figure}

\begin{figure}[!ht]
\centering  
\begin{subfigure}{0.9\textwidth}
\includegraphics[width=1\linewidth]{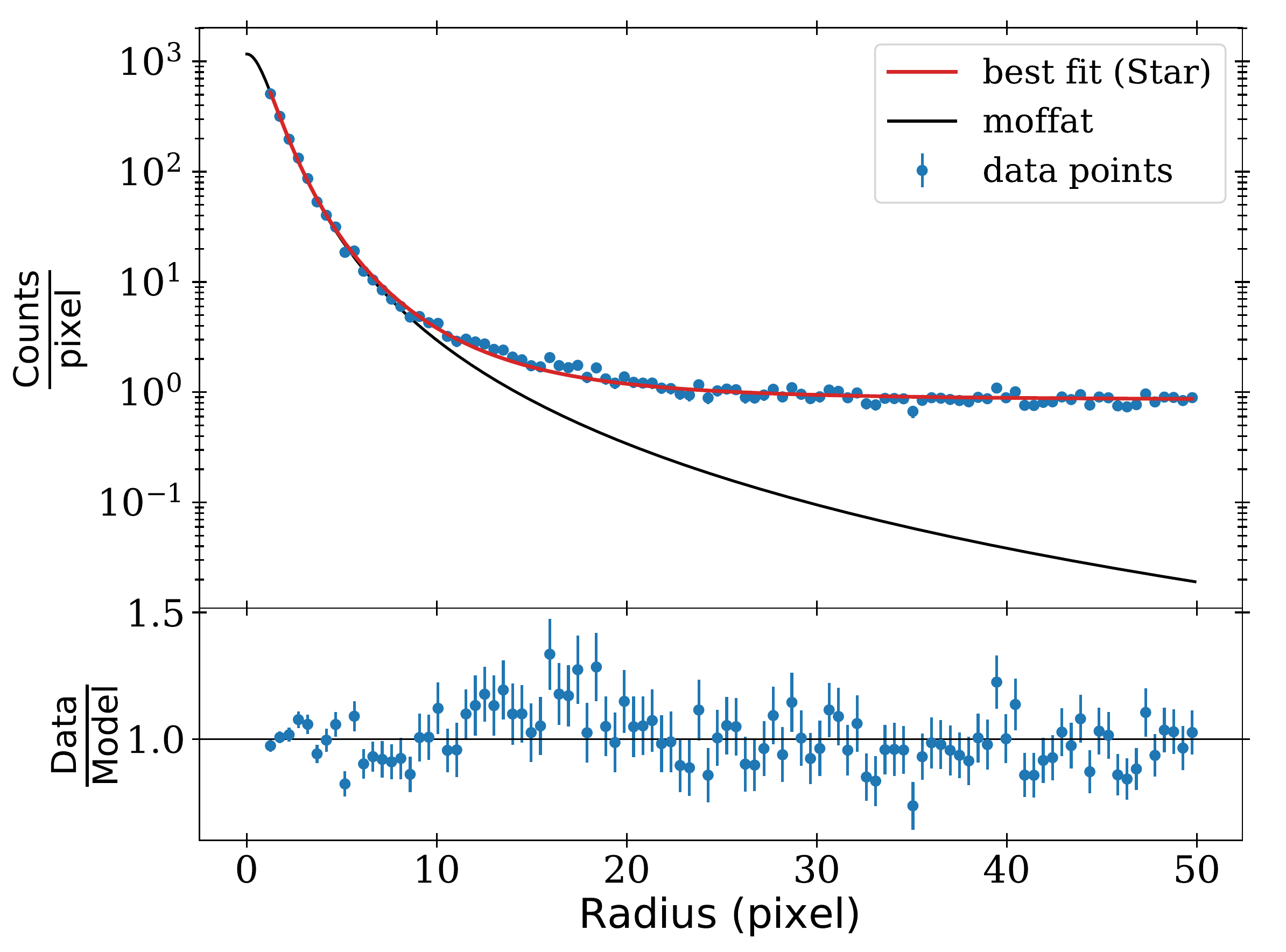}
\caption{}
\label{star_fuv}
\end{subfigure}
\medskip
\begin{subfigure}{0.9\textwidth}
\includegraphics[width=1\linewidth]{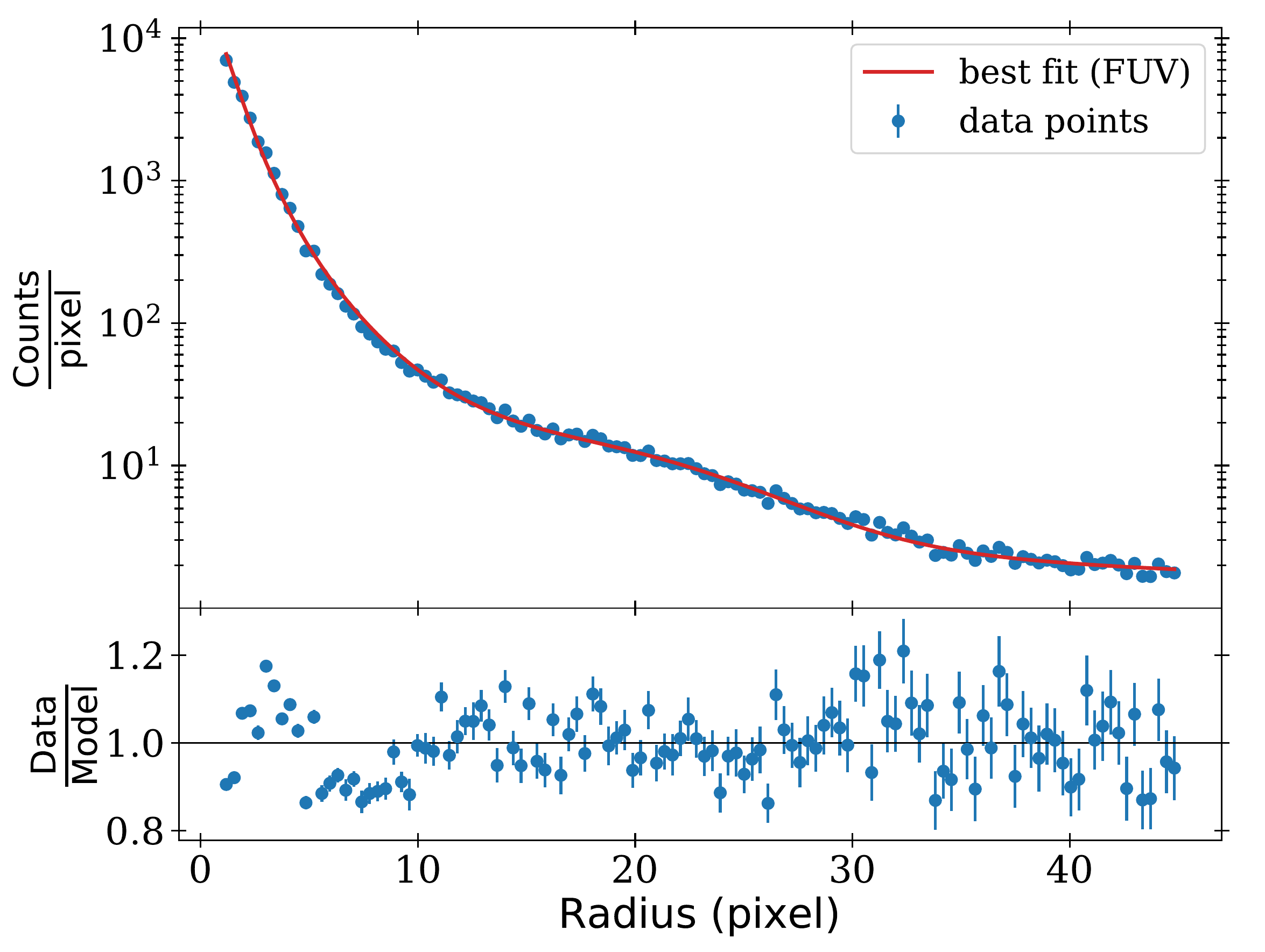}
\caption{}
\label{radial_fuv}
\end{subfigure}
\caption{\small{Radial profile fit of: (a) bright star BPS~CS~29521--0036 in FUV band, using a Moffat function and a constant background  (b) NGC~7469 in FUV band. }}
\label{radprofile_fuv}
\end{figure}

To separate the host galaxy contribution from the UVIT light curves, we fitted the radial profile of the source NGC~7469. For NUV, we first fitted the radial profile of the bright star TYC~1160-1473-1 to determine the PSF of the instrument. We fitted the core of the radial profile with the Moffat function, $M(x) = A_m \left[ 1+ \left(\frac{x}{\alpha}\right)^2 \right]^{-\beta}$  $ \left(FWHM =2 \alpha \sqrt{2^{\frac{1}{\beta}}-1}\right)$ and the extended wings with the exponential function, $I(x) = I_0\times exp^{\left(\frac{-x}{d}\right)}$ plus a constant background (Fig.~\ref{star_nuv}). The star TYC 1160-1473-1 was not bright enough in the FUV band, so we used the bright star BPS~CS~29521--0036 to fit the PSF with a single Moffat function plus a constant for the background  (Fig.~\ref{star_fuv}). \\

The FWHM for the best fitting inner Moffat profile of the PSF for NUV (N245M) is 0.82 arcsecs and for FUV (F172M) is 0.94 arcsecs. We used the best fit PSF profile with variable amplitude of the Moffat Function to fit the radial profile of the central part of NGC~7469, which should represent the AGN flux profile.  For the host galaxy distribution, we used an additional exponential function  and a gaussian model $\left(f(x) = A_G \times exp\left[ \frac{-4log(2)(x-x_0)^2}{fwhm^2}\right] \right)$. \\


Fig.~\ref{radial_nuv}, \ref{radial_fuv} shows the AGN plus the galaxy profiles in the NUV and FUV bands, respectively. Statistically speaking, the fits are not acceptable (large $\chi^2$ value) but, as these figures show, the  model fits capture well the overall radial light distribution in both bands. Using the best-fit model parameters, we found that the host galaxy contribution is $\sim 3.5$ \% in FUV and $\sim 5$ \% in NUV, within 18 pixels  radius (this is the size of the aperture we used to extract the light curve from the central region in this source).



\bsp	
\label{lastpage}

\end{document}